\documentclass[]{cit_lab_mfr}

\usepackage{graphicx}
\usepackage{subcaption}
\usepackage{wrapfig}
\usepackage{tikz}
\usetikzlibrary{positioning,decorations.pathreplacing,calc,arrows.meta,fit,backgrounds}
\usepackage{algorithm}
\usepackage{algpseudocode}
\usepackage{xspace}
\usepackage[toc,page,header]{appendix}

\graphicspath{{./figures/}}

\newcommand{\paratitle}[1]{\par\addvspace{1.5ex}\noindent\textbf{#1}}
\providecommand{\ie}{}\renewcommand{\ie}{\emph{i.e.,}\xspace}
\providecommand{\eg}{}\renewcommand{\eg}{\emph{e.g.,}\xspace}

\newcommand{\Description}[1]{}

\newcommand{\modelname}{OneTrans-V2}   %
\newcommand{\Model}{\modelname\xspace}
\newcommand{\dgrfull}{Decision-Conditioned Generative Retrieval\xspace}  %
\newcommand{\dgr}{DCGR\xspace}

\hypersetup{pdftitle={OneTrans-V2: Unifying Retrieval, Pre-rank, and Fine-rank with One Transformer in Industrial Recommender}, pdfauthor={ByteDance Global E-Commerce Recommendation Foundation Team}}
\title{OneTrans-V2: Unifying Retrieval, Pre-rank, and Fine-rank with One Transformer in Industrial Recommender}

\affiliation{%
\parbox{\textwidth}{\centering
\textbf{ByteDance Global E-Commerce Recommendation Foundation Team}
}}

\renewcommand{\thefootnote}{\fnsymbol{footnote}}
\renewcommand{\thefootnote}{\arabic{footnote}}  %

\abstract{
Industrial recommendation systems typically operate as a \emph{cascade} of retrieval, pre-rank, and fine-rank, but these stages are usually trained and served as separate models, causing repeated user-sequence encoding, isolated optimization, and duplicated engineering effort. Building on OneTrans' model-level unification, we present OneTrans-V2, one Transformer that unifies the entire cascade. It encodes the user behavior sequence once as a shared context while preserving stage-specific candidate features and computation. Joint training lets the three stages reinforce one another and enables in-model knowledge distillation from fine-rank to pre-rank. We scale the shared backbone with sparse mixture-of-experts (MoE), which increases capacity with bounded activated computation, and stabilize scaling with $\mu$P-style parameterization. To consolidate objective-specific retrieval channels, we introduce Decision-Conditioned Generative Retrieval (DCGR). DCGR predicts a decision prefix describing the upcoming interaction and generates items conditioned on it, allowing business objectives to steer a single generative process. Finally, Sequence-Native Training (SNT) organizes training around each user's lifelong behavior sequence and amortizes its encoding across exposures. Deployed across all three stages of a large-scale industrial recommendation system, OneTrans-V2 improves gross merchandise value (GMV) by 9.74\% and, with a co-designed serving stack, delivers $3.2\times$ the throughput of the cascade it replaces under the same hardware budget.
}

\date{\today}

\begin{document}
\maketitle

\begingroup
\renewcommand{\thefootnote}{\fnsymbol{footnote}}
\endgroup

\section{Introduction}\label{sec:intro}

Modern industrial recommendation systems operate as a \emph{cascade} to meet stringent latency constraints. Retrieval selects thousands of candidates from a billion-scale catalog, pre-rank reduces them to hundreds with inexpensive per-candidate scoring, and fine-rank produces the final order using richer candidate features, including user--item cross features. This progressive reduction makes billion-scale recommendation feasible within a few hundred milliseconds.

\begin{figure}[!t]
\centering
\includegraphics[width=0.7\linewidth]{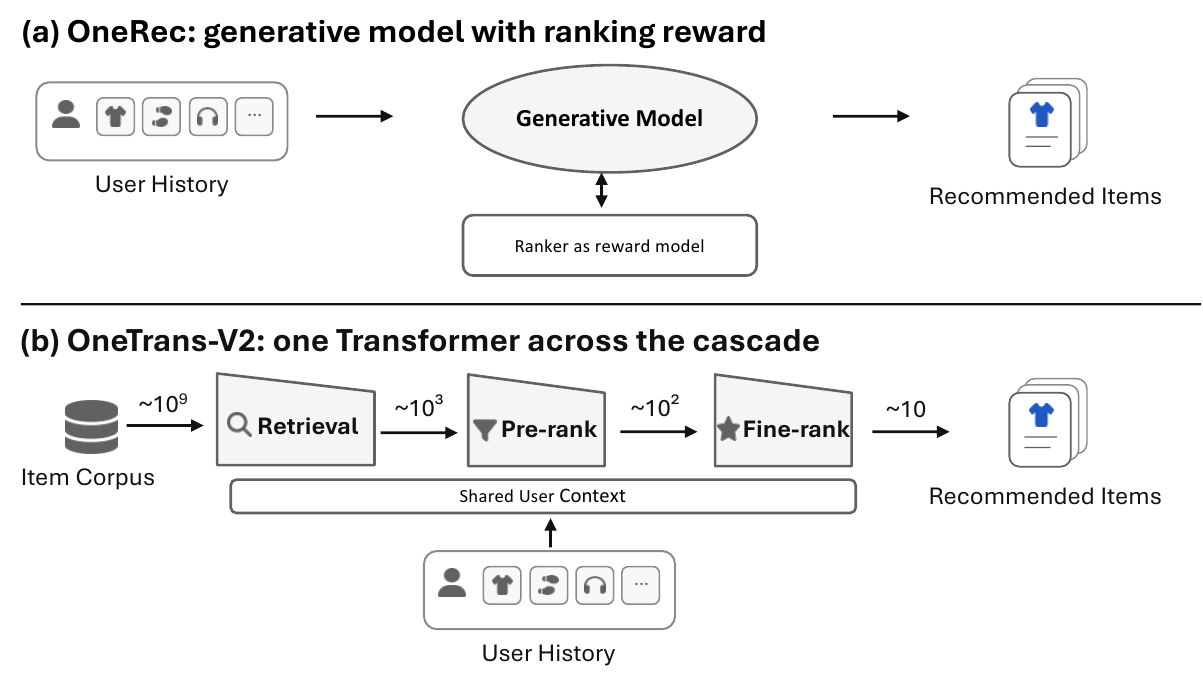}
\caption{Two routes to unifying an industrial recommendation cascade. (a): End-to-end generative recommendation, exemplified by OneRec, replaces the cascade with one generative model and uses a ranker as its reward model. (b): \Model keeps the retrieval, pre-rank, and fine-rank cascade and its candidate-set reduction, but runs the three stages as tasks of one Transformer over a shared, cached user context.}
\vspace{-5mm}
\label{fig:overall}
\end{figure}

In conventional cascades, however, retrieval, pre-rank, and fine-rank are designed, trained, and served independently, and retrieval may further maintain multiple models for different business objectives. This fragmentation is costly in four ways. 1) Engineering effort is duplicated across stages. A new user feature, a longer behavior sequence, or a model-scaling improvement must be implemented, tuned, validated, and deployed separately for each stage. 2) The same user behavior sequence is repeatedly encoded by every stage, wasting substantial computation. 3) Stages optimized in isolation cannot directly share learned representations or supervision, which can lead to inconsistencies across the cascade. 4) Model capacity is fragmented across specialized models, limiting the ability to scale the cascade as a whole.

Individual ranking models once suffered from a similar fragmentation. They traditionally relied on specialized modules for feature interaction and behavior modeling. Our previous work, OneTrans~\cite{onetrans}, showed that these components can be unified within a single causal Transformer. However, this \emph{model-level} unification remains confined to a single ranking stage, leaving retrieval, pre-rank, and fine-rank as separate models. An alternative route, represented by OneRec~\cite{onerec}, eliminates the conventional cascade with an end-to-end generative model (Figure~\ref{fig:overall}(a)). While this provides a direct path to architectural unification, industrial ranking stages rely heavily on rich candidate features. Removing these features can substantially degrade recommendation quality~\cite{mtgr}. We therefore pursue a \emph{cascade-preserving} approach (Figure~\ref{fig:overall}(b)). Rather than eliminating the cascade, we retain each stage's native candidate modeling while unifying the candidate-independent user context shared across stages.

Building on OneTrans, we present \Model. It extends Transformer-based unification from a single ranking model to the whole recommendation cascade by representing retrieval, pre-rank, and fine-rank as three tasks of a single causal Transformer. Specifically, in \Model the candidate-independent user behavior sequence is encoded into a shared user context whose states are reused across stages. Each stage appends its own tokens to this context and uses token-specific parameters under OneTrans's mixed parameterization to model its heterogeneous features. The behavior sequence is therefore encoded once per request rather than once per stage. The three stages are also trained jointly, allowing them to improve the shared user context rather than being optimized in isolation. A stage visibility mask (\S\ref{ssec:one-transformer}) controls how stage-specific tokens access the shared user context. Each stage can read the shared user context, but its tokens remain isolated from tokens of other stages and cannot modify it. This isolation allows each stage to choose its token budget independently according to its latency and feature requirements. Pre-rank remains lightweight with a single token over a small feature set, whereas fine-rank uses multiple tokens to model its richer candidate features.

Beyond computation reuse, unification pools the capacity previously dispersed across specialized models into a single backbone that can be scaled as a whole. Dense scaling, however, increases model capacity together with computation and memory traffic~\cite{routedscaling,ttinference}. We therefore scale the pooled backbone with sparse mixture-of-experts (MoE)~\cite{moe,switch}, decoupling total model capacity from activated computation and per-token memory traffic. As the backbone grows, we further adopt $\mu$P-style parameterization to stabilize optimization across model scales. Unification also simplifies cross-stage knowledge distillation. Pre-rank and fine-rank are conventionally optimized as separate models and can produce inconsistent candidate orderings~\cite{cold,copr}. Existing distillation methods~\cite{rankdistill,rankflow,corr} address this inconsistency but require a separate fine-rank teacher to be trained, scored, and refreshed independently. In \Model, the two ranking stages are tasks of the same Transformer and therefore share the user context and update in the same training step. Fine-rank can therefore directly supervise pre-rank within the model, improving consistency between the two stages.

Retrieval, however, poses a distinct challenge to unification. In our production system, different business objectives call for different candidate sets. The main channel targets clicks and orders, while dedicated channels promote discovery or advertising. Retrieval has therefore evolved into multiple objective-specific channels, each with its own model and decoding process. The difficulty is that these channels differ not merely in what they predict, but in what they retrieve. In pre-rank and fine-rank, candidates are already given, so multiple objectives can share one forward pass through separate prediction heads. In retrieval, the objective determines which candidates are generated in the first place. Keeping a separate decoding process for each objective would therefore preserve the very fragmentation we seek to remove. A unified retriever must instead express multiple business objectives within a single generation process.

We resolve this with \dgrfull (\dgr). Instead of generating an item directly from the user context, \dgr introduces an intermediate decision step before item generation. The model first predicts a decision prefix that represents a user-conditional distribution over the upcoming interaction. The decision dimensions capture objective-related outcomes, such as whether the user converts, at what value, and how new the category is to the user. The model then generates the item's semantic identifier (SID) conditioned on this prefix. In this way, decision-conditioned generation provides a compact, non-linguistic analogue of chain-of-thought reasoning. The model represents intermediate decisions as decision tokens rather than in natural language and uses them to guide subsequent item generation.
This decision space also provides a common interface for business objectives. Instead of assigning each objective its own retrieval channel, we apply a business offset to steer the predicted decision distribution toward the desired objective. The offset changes how decision prefixes are prioritized while leaving the item-generation model shared across objectives. Its strength is exposed as an online parameter. The same model can therefore trade off different business objectives without retraining or adding a new retrieval channel.

Once the three stages share one backbone, encoding the long user behavior sequence becomes its dominant training cost. We therefore introduce Sequence-Native Training (SNT). SNT organizes exposures around each user's lifelong behavior sequence and amortizes sequence encoding across the user's exposures. Each exposure retains its own causal view of the shared behavior sequence while reusing the same user-side computation. This yields a 4.4$\times$ training speedup.

We deploy \Model across retrieval, pre-rank, and fine-rank in a large-scale industrial recommendation system. It improves gross merchandise value (GMV) by 9.74\%. With a co-designed serving stack, \Model delivers \(3.2\times\) the queries per second (QPS) of the cascade it replaces under the same hardware budget.

In summary, our contributions are as follows:
\begin{itemize}

\item \textbf{Cascade-level One Transformer.} We formulate retrieval, pre-rank, and fine-rank as three tasks of one jointly trained Transformer, reusing the candidate-independent user context across stages while preserving each stage's native candidate features, heterogeneous objectives, and serving budgets. This unification enables computation reuse, cross-stage knowledge transfer, and capacity scaling of the whole cascade within a single backbone. We show empirically that joint training improves all three stages over their separately trained counterparts, while distillation within the model improves consistency between pre-rank and fine-rank.

\item \textbf{Objective consolidation in retrieval.} In our production system, retrieval remains split into objective-specific channels because it generates rather than scores candidates. We introduce \dgr, which predicts a user-conditional decision distribution from the behavior sequence and conditions identifier generation on the resulting decision prefix. Business objectives steer that distribution through a business offset, so one generative model covers multiple objectives without an external routing rule.

\item \textbf{Production-efficient training, scaling, and validation.} We use sparse MoE to scale model capacity with bounded activated computation, adopt $\mu$P-style parameterization to stabilize optimization across model scales, and introduce SNT to keep the unified backbone within a production training budget. We validate \Model in a large-scale industrial system, where it improves GMV by 9.74\%. With a co-designed serving stack, \Model delivers \(3.2\times\) the end-to-end QPS of the cascade it replaces.

\end{itemize}
\section{Related Work}

\subsection{Model-Level Unification in Ranking}
Industrial ranking models traditionally combine specialized components for feature interaction and user behavior modeling, including feature-interaction architectures such as DeepFM~\cite{deepfm} and DCN V2~\cite{dcnv2}, and behavior encoders such as DIN~\cite{din} and SIM~\cite{sim}.
Recent work increasingly integrates these functions within unified token-based or shared-backbone architectures.
RankMixer~\cite{rankmixer} emphasizes hardware-efficient feature interaction through token mixing, while MTGR~\cite{mtgr} incorporates rich candidate features into scalable sequence modeling.
TokenFormer~\cite{tokenformer} and OneTrans~\cite{onetrans} more explicitly unify heterogeneous feature interaction and user behavior modeling within a single ranking model.
Together, these works demonstrate the value of reducing architectural fragmentation within an individual ranking stage.
\Model extends this direction from \emph{model-level} unification within a stage to \emph{cascade-level} unification across retrieval, pre-rank, and fine-rank.

\subsection{From Cross-Stage Collaboration to Cascade-Level Unification}
Cross-stage methods have progressively evolved from transferring information between independently optimized models toward directly sharing representations and computation across stages.
Early approaches such as RankFlow~\cite{rankflow} and cooperative retriever and ranker (CoRR)~\cite{corr} improve cross-stage consistency through supervision propagation, adaptive sampling, and distillation, while largely retaining separate parameters, user-side computation, and serving paths.

OneRec~\cite{onerec} represents one route toward unifying the cascade.
It reformulates recommendation around generative retrieval, where ranking is absorbed into the generation process through ranking-oriented supervision or policy signals rather than served as a separate online stage.
Concurrent work such as Sona~\cite{sona} further extends this direction toward retrieval, pre-rank, and fine-rank within a unified encoder--decoder architecture.
Meanwhile, feature-rich industrial rankers such as MTGR~\cite{mtgr} continue to rely on stage-specific candidate features, motivating designs that keep such heterogeneous features while sharing reusable computation.

A second route preserves the cascade while more tightly coupling retrieval and ranking.
UniPinRec~\cite{unipinrec} jointly designs retrieval and lightweight ranking across representation, training, and serving, while OneRanker~\cite{oneranker} and UniSGR~\cite{unisgr} connect generative retrieval with downstream ranking through shared representations and ranking-oriented supervision.
UniR$^2$~\cite{unir2} further places user context, the SID generation trajectory, and ranking-side item feature tokens in a single heterogeneous decoder-only sequence.
Its ranking query directly attends to the generation trajectory, which serves as a representation bridge from retrieval to ranking. Shared base attention parameters, ranking-specific FFNs, and low-rank adaptation (LoRA) balance representation sharing with task adaptation and optimization isolation.

\Model takes this second route but couples the stages through the shared user context and through a distillation loss between its two ranking stages.
Unlike UniR$^2$, whose ranking query attends to the generation trajectory, the stage-specific tokens of \Model remain isolated across stages in the forward pass rather than consuming intermediate representations produced by another stage.

\subsection{Decision-Conditioned Generative Retrieval}
Generative retrieval represents items with semantic identifiers and autoregressively generates candidates conditioned on user context, as exemplified by TIGER~\cite{tiger} and OneRec~\cite{onerec}. 
Recent work further incorporates objective-related signals into generation. PinRec~\cite{pinrec} conditions retrieval on desired outcome representations, while UniSGR~\cite{unisgr} prepends learned task-aware tokens for click, add-to-cart, and purchase.

\Model differs in that \dgr does not take the objective as an input token. 
Instead, it predicts the decision dimensions from the user context and uses the inferred decision prefix to condition subsequent SID generation. 
In this way, \dgr jointly learns decision inference and decision-conditioned retrieval, allowing one generative model to serve multiple business objectives.

Steering what a generative model recommends has two approaches. One attaches a reward to what the model generates and optimizes it with reinforcement learning, as OneRec~\cite{onerec} does with a ranker serving as the reward model. The other writes the outcome into the sequence itself and trains the model by ordinary next-token prediction. Decision Transformer~\cite{decisiontransformer} and the outcome-conditioned methods that follow it~\cite{udrl,gcsl} take this second approach. They place the outcome of a sequence of actions in front of those actions, so the desired outcome can be written in at inference and the actions follow from it.

\dgr is the single-step case of that idea. A decision describes the one interaction at hand rather than the total reward collected over a long sequence of actions, so its labels are read directly from the log and nothing has to be estimated. Two costs of the first approach disappear with it. No reward model has to score what the model generates, and the logged samples need not be reweighted to account for having been produced by an earlier model rather than by the one being trained. Being single-step also addresses a known limitation of outcome conditioning. Given only sequences of actions that are individually poor, it cannot assemble a better one out of their good parts, as methods that plan several steps ahead can. \dgr predicts one decision and then the item, so there is no sequence to assemble and the limitation does not apply.

The closest prior method at inference is Multi-Game Decision Transformer~\cite{mgdt}, which does not need the desired outcome to be supplied either. It predicts a distribution over outcomes, shifts that distribution toward the better ones, and then generates an action. The business offset of \S\ref{sssec:dig} is that shift, applied to structured decisions rather than to a single number. That model draws one outcome from the shifted distribution and conditions on it, whereas \dgr scores every decision prefix. Each prefix carries its shifted score into one beam search over the decision and the SID codes together. A prefix that scores well on its own but leads to no good item is therefore outranked instead of being chosen.

\subsection{Sequence-Centric Training and Compute Reuse}
Recent work increasingly reorganizes recommendation training and inference around shared sequence or request contexts to reduce repeated encoding of the user behavior sequence. 
HSTU~\cite{hstu} formulates recommendation as sequential transduction, and its M-FALCON inference algorithm reuses cached history states across candidate microbatches and requests. Request-level batching (RLB)~\cite{rlb}, SORT~\cite{sort}, and MTGR~\cite{mtgr} amortize user-side computation by grouping the candidates of a request or a user around one shared context.

SNT applies this reuse principle to the unified cascade. Within each user window, exposures from different requests and cascade stages share one encoding of the user's behavior sequence (\S\ref{ssec:snt}).

\section{Background}
\label{sec:background}
OneTrans~\cite{onetrans} is a Transformer-based ranking model that jointly models sequential user behaviors and heterogeneous non-sequential features. It projects each behavior event into a sequential token, forming a behavior sequence $\mathcal{S}$, while non-sequential features are converted into a set of tokens $\mathcal{NS}$ through a dedicated tokenizer.

The key design of OneTrans is its \emph{mixed parameterization}. Sequential tokens follow a homogeneous representation scheme and therefore share the same attention ($Q/K/V$) and feed-forward (FFN) parameters, whereas heterogeneous non-sequential tokens use \emph{non-shared} (token-specific) $Q/K/V$ and FFN parameters. A causal attention mask places the non-sequential tokens after the behavior sequence, allowing them to attend to the entire preceding behavior sequence without violating causality. The resulting non-sequential token states are then fed into task heads to produce ranking predictions.

This decomposition separates candidate-independent behavior-sequence modeling from heterogeneous non-sequential feature interaction, providing the architectural basis for \Model, since the former can be shared across cascade stages while the latter remains stage-specific.

\section{OneTrans-V2}\label{sec:model}
\begin{figure*}[t]
\centering
\includegraphics[width=\textwidth]{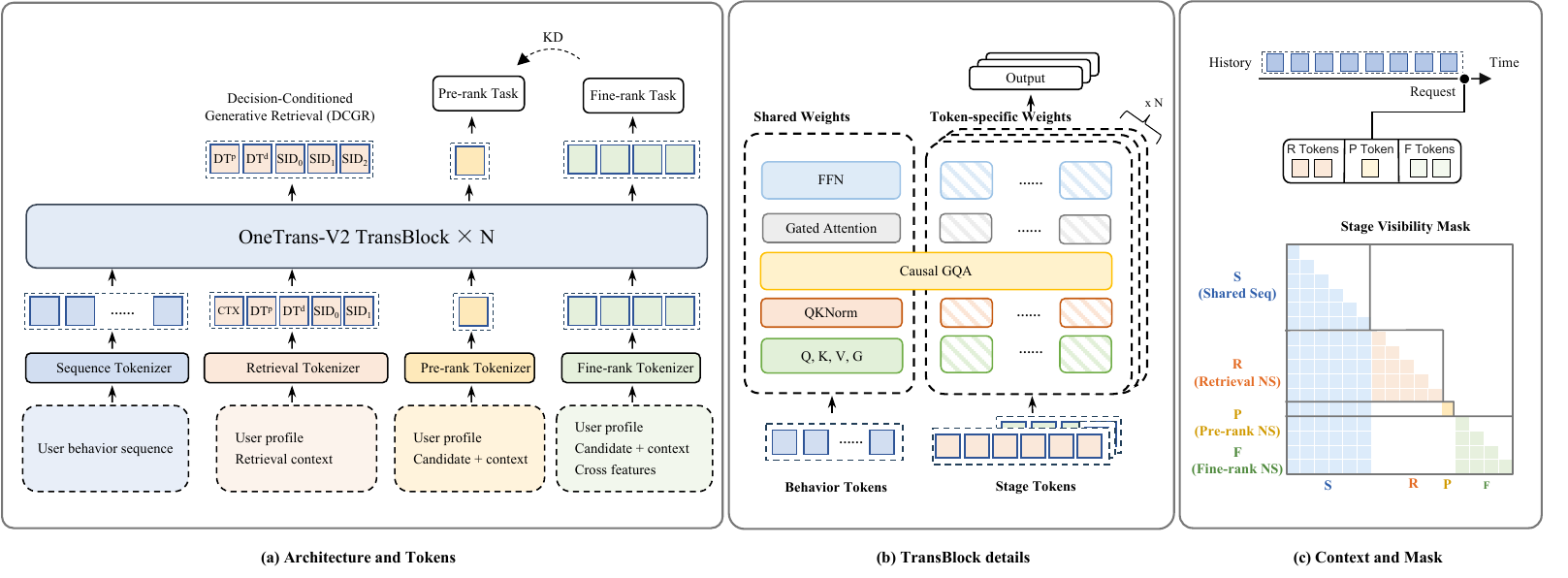}
\caption{Architecture of \Model, with $N$ Transformer blocks.
(a) Shared user behavior tokens and stage-specific non-sequential (NS) tokens are produced by separate tokenizers and processed by the unified backbone.
\dgr uses the retrieval context token \texttt{CTX}, parallel and dependent decision tokens $\texttt{DT}^{p}$ and $\texttt{DT}^{d}$, and three SID codes $\texttt{SID}_0$--$\texttt{SID}_2$ to generate items.
Pre-rank and fine-rank heads score candidates; the dashed KD arrow denotes knowledge distillation from the fine-rank teacher to the pre-rank student.
(b) Behavior tokens share parameters, whereas stage-specific tokens use parameters specific to each token.
$Q$, $K$, and $V$ denote query, key, and value projections. $G$ denotes the gate of gated attention.
(c) $S$ denotes the shared behavior sequence, and $R$, $P$, and $F$ denote the retrieval, pre-rank, and fine-rank tokens, respectively.
The request timestamp anchors the tokens of each stage to its visible behavior prefix.
In the mask, rows are queries and columns are keys; colored cells permit attention and white cells mask it.
Behavior tokens attend causally within $S$. Each stage-specific token attends to its visible behavior prefix, itself, and preceding tokens in the same stage and exposure; tokens from other stages or exposures are masked.}
\label{fig:model}
\end{figure*}

Conventional cascades cut along the stage boundary, so every stage has its own model and re-encodes the common user behavior sequence for itself. \Model (Figure~\ref{fig:model}) instead cuts between what repeats and what is stage-specific, encoding that sequence once into a single user context and letting every stage append its own tokens to it, so each stage keeps its native candidate-side modeling. \Model therefore preserves the cascade and its stage-specific modeling. Yet it needs only one jointly trained causal Transformer, whereas the conventional design needs a separate model for every stage and, within retrieval, for every business objective.

\subsection{One Transformer across the Cascade}
\label{ssec:one-transformer}

The inputs to \Model fall into two parts. The candidate-independent user behavior sequence $\mathcal{S}$ is shared across the cascade. The non-sequential feature sets $\mathcal{NS}_r$, $\mathcal{NS}_p$, and $\mathcal{NS}_f$ are specific to retrieval, pre-rank, and fine-rank; each is tokenized only by its own stage into tokens that follow the shared user context. Following OneTrans (\S\ref{sec:background}), \Model applies \emph{mixed parameterization} to the two token types within every Transformer block (TransBlock). The behavior tokens in $\mathcal{S}$ share one set of attention ($Q/K/V$) and FFN parameters, whereas the non-sequential tokens of each stage, which we call \emph{stage-specific tokens}, use their own non-shared, token-specific $Q/K/V$ and FFN parameters. This allows all three stages to operate within one Transformer while preserving their heterogeneous feature representations; the internal design of each block is given in \S\ref{ssec:scaling}.

\paratitle{Shared Causal User Context.} The user behavior sequence $\mathcal{S}=(s_1,\ldots,s_L)$ of a user $u$, also written $\mathcal{S}_u$, lists behavior events in the order in which they arrive, with each new event appended to the sequence. To avoid future-information leakage, we follow OneTrans~\cite{onetrans} and model $\mathcal{S}$ with a causal attention block, so that the hidden state at each position summarizes only the behaviors that precede it in the sequence.

\paratitle{Stage Visibility Mask.}
A \emph{request} is one recommendation call from a user, served by all three stages in turn. The stage-specific tokens of an exposure belong to a request made at some timestamp. We \emph{anchor} them at the last behavior that had arrived by that timestamp, so a stage-specific token attends to the behavior prefix up to its anchor and not to any behavior that arrives later. Within the same exposure and stage, stage-specific tokens further follow causal attention according to their predefined token order. Tokens from different stages or different exposures are mutually masked, and behavior tokens never attend to stage-specific tokens, so the user context can be encoded once and read by every stage. Although stage-specific tokens are isolated from one another, all three tasks jointly optimize the shared user context, allowing retrieval, pre-rank, and fine-rank to share user modeling without mixing their non-sequential features. How exposures from different requests share one encoded sequence under this mask is the subject of \S\ref{ssec:snt}.

\subsection{Task Formulations and Joint Optimization}
\label{ssec:objectives}

Given the shared user context and stage-specific tokens described above, \Model jointly optimizes three heterogeneous tasks, namely autoregressive generation for retrieval and discriminative prediction for pre-rank and fine-rank.

\subsubsection{Decision-Conditioned Generative Retrieval}
\label{sssec:dig}

In generative retrieval, the model does not score given candidates. Instead, it autoregressively generates item identifiers conditioned on the user context. To merge the objective-specific channels of \S\ref{sec:intro} into one such model, we need two things. First, items need a structured identifier that can be generated. Second, the model needs an internal representation of the decision that each channel builds in, so that business objectives can steer it. Semantic identifiers provide the first, and \dgrfull (\dgr) provides the second.

\paratitle{Semantic Identifiers.}
Following prior generative retrieval work~\cite{tiger,onerec}, we represent each item with a three-level semantic identifier (SID) $(\texttt{SID}_0,\texttt{SID}_1,\texttt{SID}_2)$ obtained by applying residual-quantized k-means (RQ-KMeans)~\cite{qarm} to its multimodal embedding. The three codes form a coarse-to-fine hierarchy, and the model generates them autoregressively in that order. Retrieval is implemented as a beam search of width $k$ over this code tree rather than as a scoring pass over a fixed candidate set. Because multiple items may share the same codes, every item mapped to a retrieved SID is returned.

\paratitle{Decision-Conditioned Generative Retrieval.}
Plain generative retrieval goes straight from the user context to the SID codes, with nothing observable in between. \dgr inserts an intermediate step. Before generating the SID codes, the model first predicts a few \emph{decisions} about the upcoming interaction, namely its \emph{purchase level}, \emph{spending level}, \emph{discovery level}, and \emph{supply type} (whether the user will buy, at what value, how new the category is to the user, and whether the item is sponsored). These predictions form a \emph{decision prefix}, a few decision tokens placed before the SID codes. Chain-of-thought prompting~\cite{wei2023chainofthoughtpromptingelicitsreasoning} shows how much a language model gains from intermediate steps, and \dgr adds such steps to retrieval, supervised by the log rather than generated freely. A \emph{decision} here is an outcome that the model predicts, not a choice that the user makes explicitly. The four dimensions span a \emph{decision space}, and each business objective corresponds to a region of this space rather than to a separate retrieval channel.

Consider a single request. A user arrives, and before generating any item the model predicts what kind of interaction is coming. It might predict a purchase at the middle spending level, a high discovery level, and organic supply. It then decodes the SID codes conditioned on that prediction, so the items it returns are the ones that fit an interaction of that shape. Had it predicted no purchase at a low discovery level, the same user context would have produced different items. This is why one model can serve several objectives. The objective is not a routing decision made outside the model, but a region of the prefix the model itself predicts, and a business offset moves the prediction toward that region.

Decisions come in two types. \emph{Parallel decisions} depend only on the user context and are predicted at the same position. \emph{Dependent decisions} also depend on earlier decisions and therefore take a later, \emph{chained} position.
The retrieval-specific features $\mathcal{NS}_r$ are summarized into a context token \texttt{CTX} that follows the shared behavior sequence $\mathcal{S}$. Conditioned on $\mathcal{H}=(\mathcal{S},\texttt{CTX})$, the model generates from the \texttt{CTX} position onward, one position after another,
\begin{equation}
[\,\underbrace{\texttt{DT}^{p}}_{\text{parallel decisions}},\;
\underbrace{\texttt{DT}^{d}}_{\text{dependent decisions}},\;
\underbrace{\texttt{SID}_0,\texttt{SID}_1,\texttt{SID}_2}_{\text{item}}\,].
\label{eq:dig-seq}
\end{equation}

The four decision dimensions take their labels directly from the behavior logs.

\emph{Parallel decisions.}
Three dimensions can be predicted from the user context alone.
\emph{Purchase level} ($z_{oc}$, short for order count) says whether the interaction ends in no purchase, one purchase, or several. Because $z_{oc}{=}0$ already marks a non-converting interaction, no separate purchase indicator is needed.
\emph{Discovery level} ($z_{disc}$, short for discovery) says how new the candidate category is relative to the user's recent behavior.
\emph{Supply type} ($z_{ad}$, short for advertising) says whether the item is organic or sponsored.
Each of the three is predicted from the hidden state at the \texttt{CTX} position by its own lightweight decision head. Their decision embeddings are summed into one parallel decision token $\texttt{DT}^{p}$, which takes the next position. We sum rather than chain them for two reasons. Chaining would force us to pick a decoding order among dimensions that have no dependence to order them by. That choice is not neutral, since whichever dimension is decoded first is served best (\S\ref{ssec:prefix-design}). Summing also keeps the decision prefix short.

\emph{Dependent decisions.}
The only dependent decision in this paper is the \emph{spending level} ($z_{aov}$, short for average order value), the value range of each order. We deliberately do not model transaction value as a single GMV label, because such a label cannot tell a basket of several cheap orders from one expensive purchase. With transaction value split into purchase level and spending level, the spending level is meaningful only once the purchase level is known. It is therefore predicted after $\texttt{DT}^{p}$ and represented by the dependent decision token $\texttt{DT}^{d}$. The SID codes are then generated one by one, conditioned on the user context and the predicted decisions.

\paratitle{\dgr Objective.}
\dgr models the joint distribution of the decisions and the item as
\begin{equation}
P(\mathbf{z},\texttt{Item}\mid\mathcal{H})
=
\underbrace{P(\mathbf{z}\mid\mathcal{H})}_{\text{Decision Modeling}}
\cdot
\underbrace{P(\texttt{Item}\mid\mathcal{H},\mathbf{z})}_{\text{Decision-Conditioned Retrieval}},
\label{eq:dig-factor}
\end{equation}

where
$\mathbf{z}^{p}=(z_{oc},z_{disc},z_{ad})$
denotes the parallel decisions and
$\mathbf{z}=(\mathbf{z}^{p},z_{aov})$
the complete decision prefix.
With the two types of decisions organized as above,
\begin{equation}
\begin{array}{@{}l@{\;}l@{}}
P(\mathbf{z}\mid\mathcal{H})
&=
\displaystyle
\prod_{j\in\{oc,disc,ad\}}
P(z_j\mid\mathcal{H})
\cdot
P(z_{aov}\mid\mathcal{H},\mathbf{z}^{p}),
\\[4pt]
P(\texttt{Item}\mid\mathcal{H},\mathbf{z})
&=
\displaystyle
\prod_{\ell=0}^{2}
P\!\left(
\texttt{SID}_\ell
\mid
\mathcal{H},\mathbf{z},\texttt{SID}_{<\ell}
\right).
\end{array}
\label{eq:dig-expand}
\end{equation}

The first line predicts the decisions from the user context. The second line generates the item given the decisions.
All decision and SID predictions are trained with cross-entropy in the autoregressive order of Eq.~\ref{eq:dig-seq}, with the ground-truth decisions supplied at their positions (teacher forcing). The resulting retrieval loss is $\mathcal{L}_r$.

\paratitle{Business Steering.}
Decoding is a beam search over the decision prefix and the SID codes, so which items are returned depends on how the decision prefixes score against one another. A business objective enters at decoding time by changing that ranking. Decoding proceeds in three steps. First, every valid combination $\mathbf{z}$ of the decisions is scored over the cached user context, which gives $\log P(\mathbf{z}\mid\mathcal{H})$ for every decision prefix (Figure~\ref{fig:dig-inference}). Second, a business offset $\beta\,\phi(\mathbf{z})$ is added to each of these scores. Third, beam search extends the prefixes into the SID codes, ranking each prefix together with its codes by
\begin{equation}
\underbrace{\log P(\mathbf{z}\mid\mathcal{H})+\beta\,\phi(\mathbf{z})}_{\text{steered decision prefix}}
\;+\;
\sum_{\ell=0}^{2}\log P\!\left(\texttt{SID}_\ell\mid\mathcal{H},\mathbf{z},\texttt{SID}_{<\ell}\right).
\label{eq:steer}
\end{equation}

Here $\phi_j$ assigns an offset to each value of dimension $j$. A positive offset on the highest spending level, for instance, steers retrieval toward more valuable orders. $\phi(\mathbf{z})$ sums these offsets over the prefix, and the single scalar $\beta$ sets how strong the whole offset is.

\definecolor{cSeq}{HTML}{CEE0EE}
\definecolor{cBlock}{HTML}{E1EBF4}
\definecolor{cRec}{HTML}{F8E8D8}
\definecolor{cRecDeep}{HTML}{F5D3B3}
\definecolor{cNavy}{HTML}{3E5380}
\definecolor{cSlate}{HTML}{323B4C}
\definecolor{cBase}{HTML}{8FAACF}
\definecolor{cBoost}{HTML}{E8924A}
\definecolor{cBoostD}{HTML}{B8601A}
\providecommand{\digscalel}{1.2}  %
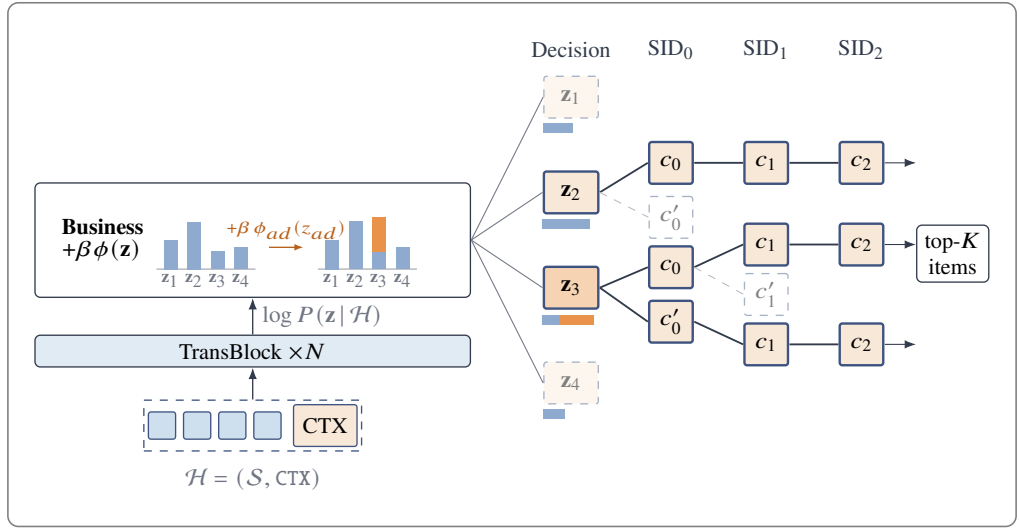
\begin{figure*}[t]
\centering
\begin{tikzpicture}[
  scale=\digscalel, every node/.append style={transform shape},
  font=\scriptsize,
  tok/.style={draw=cNavy, line width=0.5pt, rounded corners=1pt, minimum height=4.6mm,
              inner xsep=2.2pt, inner ysep=1pt, align=center, fill=cRec},
  seq/.style={tok, fill=cSeq, minimum width=3mm, minimum height=3mm, inner sep=0pt},
  dt/.style={tok, minimum width=6mm},
  dtk/.style={dt, line width=0.9pt},
  dtb/.style={dt, line width=0.9pt, fill=cRecDeep},
  dtx/.style={dt, dashed, draw=cNavy!45, fill=cRec!40, text=black!50},
  sidn/.style={tok, minimum width=4.8mm, inner xsep=1.5pt},
  sidk/.style={sidn, line width=0.9pt},
  sidx/.style={sidn, dashed, draw=cNavy!45, fill=white, text=black!50},
  grp/.style={draw=cNavy, dashed, line width=0.5pt, inner sep=1.4pt},
  blk/.style={draw=cSlate, line width=0.5pt, rounded corners=2pt, fill=cBlock, align=center},
  box/.style={draw=cSlate, line width=0.5pt, rounded corners=2pt, fill=white, align=center, inner sep=2.5pt},
  arr/.style={-{Latex[length=1.5mm]}, cSlate, line width=0.5pt},
  edge/.style={cSlate!70, line width=0.4pt},
  edgek/.style={cSlate, line width=0.7pt},
  edgex/.style={cSlate!35, dashed, line width=0.4pt},
  lbl/.style={font=\scriptsize},
]
\node[seq] (q1) at (-9.65mm,0) {};
\foreach \i in {2,...,4}{\pgfmathsetmacro{\j}{int(\i-1)}\node[seq, right=0.7mm of q\j] (q\i) {};}
\node[dt, right=1.2mm of q4, minimum width=7mm] (ctx) {CTX};
\node[grp, fit=(q1)(ctx)] (root) {};
\node[lbl, cSlate!70, anchor=north] at ($(root.south)+(0,-0.6mm)$) {$\mathcal{H}=(\mathcal{S},\texttt{CTX})$};

\node[blk, anchor=south, minimum width=48mm, inner ysep=2.5pt] (blk) at ($(root.north)+(0,3.5mm)$) {TransBlock $\times N$};
\draw[arr] (root.north) -- (blk.south);

\node[box, anchor=south, minimum width=48mm, minimum height=12.6mm] (bz) at ($(blk.north)+(0,4mm)$) {};
\node[lbl, anchor=west, align=left] at ($(bz.west)+(1.8mm,0)$) {\textbf{Business}\\[-1pt]$+\beta\phi(\mathbf{z})$};
\coordinate (h0) at ($(bz.center)+(-10.6mm,-3.2mm)$);
\draw[cSlate!50, line width=0.4pt] (h0) -- ++(10.9mm,0);
\foreach \i/\h in {0/3.2,1/5.2,2/1.9,3/2.4}{
  \fill[cBase] ($(h0)+(0.8mm+\i*2.6mm,0)$) rectangle ++(1.5mm,\h mm);
  \node[font=\tiny, cSlate!70] at ($(h0)+(1.55mm+\i*2.6mm,-1.25mm)$) {$\mathbf{z}_{\the\numexpr\i+1}$};
}
\draw[-{Latex[length=1.2mm]}, cBoostD, line width=0.5pt] ($(h0)+(12.4mm,2.4mm)$) -- node[font=\tiny, above=-1pt, cBoostD]{$+\beta\,\phi_{ad}(z_{ad})$} ++(3.8mm,0);
\coordinate (h1) at ($(h0)+(17.8mm,0)$);
\draw[cSlate!50, line width=0.4pt] (h1) -- ++(10.9mm,0);
\foreach \i/\h/\b in {0/3.2/0,1/5.2/0,2/1.9/3.8,3/2.4/0}{
  \fill[cBase] ($(h1)+(0.8mm+\i*2.6mm,0)$) rectangle ++(1.5mm,\h mm);
  \ifdim \b mm>0mm \fill[cBoost] ($(h1)+(0.8mm+\i*2.6mm,\h mm)$) rectangle ++(1.5mm,\b mm); \fi
  \node[font=\tiny, cSlate!70] at ($(h1)+(1.55mm+\i*2.6mm,-1.25mm)$) {$\mathbf{z}_{\the\numexpr\i+1}$};
}
\draw[arr] (blk.north) -- node[lbl, right=-0.5pt, cSlate!70]{$\log P(\mathbf{z}\!\mid\!\mathcal{H})$} (bz.south);

\coordinate (dx) at ($(bz.east)+(11mm,0)$);
\foreach \i/\y/\sty in {1/15.75/dtx,2/5.25/dtk,3/-5.25/dtb,4/-15.75/dtx}{
  \node[\sty] (z\i) at ($(dx)+(0,\y mm)$) {$\mathbf{z}_\i$};
  \draw[edge] (bz.east) -- (z\i.west);
}
\foreach \i/\h/\b in {1/3.2/0,2/5.2/0,3/1.9/3.8,4/2.4/0}{
  \fill[cBase] ($(z\i.south west)+(0,-0.5mm)$) rectangle ++(\h mm,-1.1mm);
  \ifdim \b mm>0mm \fill[cBoost] ($(z\i.south west)+(\h mm,-0.5mm)$) rectangle ++(\b mm,-1.1mm); \fi
}

\coordinate (sx0) at ($(dx)+(11mm,0)$);
\coordinate (sx1) at ($(dx)+(21.5mm,0)$);
\coordinate (sx2) at ($(dx)+(32mm,0)$);
\node[sidk] (a1) at ($(sx0)+(0, 8.5mm)$) {$c_0$};
\node[sidx] (a2) at ($(sx0)+(0, 2.5mm)$) {$c_0'$};
\node[sidk] (a3) at ($(sx0)+(0,-3.0mm)$) {$c_0$};
\node[sidk] (a4) at ($(sx0)+(0,-9.0mm)$) {$c_0'$};
\draw[edgek] (z2.east) -- (a1.west); \draw[edgex] (z2.east) -- (a2.west);
\draw[edgek] (z3.east) -- (a3.west); \draw[edgek] (z3.east) -- (a4.west);
\node[sidk] (b1)  at ($(sx1)+(0, 8.5mm)$) {$c_1$};
\node[sidk] (b3)  at ($(sx1)+(0,-0.5mm)$) {$c_1$};
\node[sidx] (b3x) at ($(sx1)+(0,-6.0mm)$) {$c_1'$};
\node[sidk] (b4)  at ($(sx1)+(0,-11.5mm)$) {$c_1$};
\draw[edgek] (a1.east) -- (b1.west); \draw[edgek] (a3.east) -- (b3.west);
\draw[edgex] (a3.east) -- (b3x.west); \draw[edgek] (a4.east) -- (b4.west);
\node[sidk] (c1) at ($(sx2)+(0, 8.5mm)$) {$c_2$};
\node[sidk] (c3) at ($(sx2)+(0,-0.5mm)$) {$c_2$};
\node[sidk] (c4) at ($(sx2)+(0,-11.5mm)$) {$c_2$};
\draw[edgek] (b1.east) -- (c1.west); \draw[edgek] (b3.east) -- (c3.west); \draw[edgek] (b4.east) -- (c4.west);

\foreach \p/\t in {dx/Decision,sx0/SID$_0$,sx1/SID$_1$,sx2/SID$_2$}{
  \node[lbl, cSlate] at ($(\p)+(0,21mm)$) {\t};
}

\node[box, anchor=west, align=center] (items) at ($(sx2)+(6mm,-1.5mm)$) {top-$K$\\items};
\foreach \n in {c1,c3,c4}{\draw[arr, line width=0.4pt] (\n.east) -- (items.west |- \n);}

\begin{scope}[on background layer]
  \node[draw=black!50, line width=0.6pt, rounded corners=4pt, fill=white, inner sep=3.5mm,
        fit=(current bounding box)] {};
\end{scope}
\end{tikzpicture}
\caption{\dgr at inference. The model scores every combination of the decision dimensions $\mathbf{z}$ over the cached context $\mathcal{H}$, which is the behavior sequence $\mathcal{S}$ followed by the \texttt{CTX} token. A business offset $\beta\,\phi(\mathbf{z})$ is added to these decision prefix scores only. In the steering box, the offset $\beta\,\phi_{ad}(z_{ad})$ goes to prefixes that carry the sponsored value, here $\mathbf{z}_3$ (orange). The offset raises its score and with it the chance that the beam search keeps it, while $\beta{=}0$ would leave the model's own ranking unchanged. Each prefix then carries its score into the beam search over the SID codes, which returns the top-$K$ items to pre-rank. In the tree, $c_\ell$ is a code at level $\ell$, a prime marks a competing code, and dashed nodes are pruned by the beam. Changing $\beta$ online therefore changes which decisions the model acts on, without adding a retrieval channel.}
\Description{An L-shaped layout: bottom-up stack of shared behavior sequence and CTX, TransBlock stack and a business steering box showing the decision distribution before and after the boost; from the steering box a beam-search tree fans out to the right over decision prefixes and SID codes ending in top-K items.}
\label{fig:dig-inference}
\end{figure*}

The offset is a soft preference rather than a rule. The first term of Eq.~\ref{eq:steer} corresponds to the distribution $q(\mathbf{z})\propto P(\mathbf{z}\mid\mathcal{H})\,e^{\beta\phi(\mathbf{z})}$, which is the solution of $\max_{q}\,\mathbb{E}_{q}[\phi(\mathbf{z})]-\tfrac{1}{\beta}\mathrm{KL}\!\left(q\,\|\,P(\cdot\mid\mathcal{H})\right)$, the KL-regularized improvement step used in offline reinforcement learning~\cite{awr}. In words, the offset raises the expected business value of the decisions, a KL penalty keeps them close to what the model learned from the log, and $\beta$ decides how far they may move. Multi-Game Decision Transformer~\cite{mgdt} applies the same idea to a single outcome score. Because $\phi$ is a sum over dimensions, the exponential factors as $e^{\beta\phi(\mathbf{z})}=\prod_{j}e^{\beta\phi_j(z_j)}$, one factor per objective. Each business objective therefore receives its own offset on its own dimension instead of one global weight.

Three things follow. First, the offset applies only to the decision part of Eq.~\ref{eq:steer}. It changes which decisions win, while the item term $P(\texttt{Item}\mid\mathcal{H},\mathbf{z})$ of Eq.~\ref{eq:dig-factor} is left alone, so under a given decision the model still returns the items it finds most likely. Second, $\beta{=}0$ gives the model's own ranking, and a larger $\beta$ moves further away from it, so the offset adjusts that ranking instead of overriding it. Third, $\beta$ is set online. Steering toward an objective that the existing dimensions already express therefore needs no retraining. A new objective becomes a new decision dimension, which adds one lightweight decision head to the same model instead of a new retrieval channel with its own model. If the new dimension has no conditional dependence, it is summed into $\texttt{DT}^{p}$ and the decision prefix does not get longer.

\paratitle{Guidance on the Item Side.}
The business offset decides which decision wins. The same design leaves room for a second control, which would decide how strongly the generated items follow that decision. Classifier-free guidance~\cite{cfg}, a technique first developed for diffusion models and later applied to language models~\cite{llmcfg}, can provide it. The idea is to score each SID code twice with the same model, once with the decision prefix and once without it, and to treat the difference between the two scores as the effect of the decision.

To make the second score possible, the decision prefix is replaced with a special null token $\varnothing$ on a small fraction of training samples, so that the model also learns to predict the SID codes when it is told nothing about the decision. At decoding time, each code would then be scored by
\begin{equation}
\log P(\texttt{SID}_\ell\mid\mathcal{H},\varnothing,\texttt{SID}_{<\ell})
+w\,\big[\log P(\texttt{SID}_\ell\mid\mathcal{H},\mathbf{z},\texttt{SID}_{<\ell})
-\log P(\texttt{SID}_\ell\mid\mathcal{H},\varnothing,\texttt{SID}_{<\ell})\big].
\label{eq:cfg}
\end{equation}
The bracket measures how much knowing the decision $\mathbf{z}$ changes the score of a code, and the weight $w$ scales that change. With $w{=}1$ the score equals the ordinary item term of Eq.~\ref{eq:steer}. With $w{>}1$ the effect of the decision is amplified, so the retrieved items follow it more closely. With $w{<}1$ the effect is weakened, and with $w{=}0$ the decision is ignored. Because both scores come from the same model, their difference is largely unaffected by how well either score is calibrated. Computing the second score requires a second pass of the decoder, but both passes read the same cached user context and differ only at the prefix and SID positions, so the extra work is confined to the beam search. Steering would then have two online parameters. The offset $\beta$ decides which decision is made, and the weight $w$ decides how closely the items follow it. The interventions of \S\ref{ssec:decision-ablation} suggest that the decoder responds to its prefix strongly enough for $w$ to have an effect. This extension is not yet deployed, and we leave its evaluation to future work.

\subsubsection{Discriminative Ranking Tasks}
\label{sssec:ranking}

\paratitle{Pre-rank.} Pre-rank scores every candidate that retrieval returns, so its per-candidate budget is the tightest in the cascade, and it can neither read the full non-sequential feature set nor afford many tokens for each candidate. We therefore restrict pre-rank to a lightweight feature subset $\mathcal{NS}_p\subset\mathcal{NS}_f$ and let its tokenizer compress the whole subset into a single candidate token. This token attends to the shared user context under the stage visibility mask. The pre-rank heads read out its resulting state and predict the pre-rank target set $\mathcal{T}_p$ of CTR and CVR, each optimized by a binary cross-entropy loss $\mathcal{L}_{p}$,
\begin{equation}
\mathcal{L}_{p} = -\sum_{t\in\mathcal{T}_p}\left[y_p^t \log \hat{y}_p^t+(1-y_p^t)\log(1-\hat{y}_p^t)\right],
\label{eq:loss-p}
\end{equation}
where $y_p^t\in\{0,1\}$ and $\hat{y}_p^t$ denote the label and the predicted probability of target $t$, respectively. 

\paratitle{Fine-rank.} Fine-rank scores only the candidates that survive pre-rank and determines the final order. Its main constraint therefore moves from per-candidate throughput to ranking fidelity, and pooling its rich candidate features into one token would erase the feature heterogeneity that this fidelity depends on. Following OneTrans~\cite{onetrans}, we let fine-rank consume the full non-sequential set $\mathcal{NS}_f$ and keep it as multiple tokens rather than the single token that pre-rank uses. These tokens attend to the shared user context under the same stage visibility mask, and their resulting states are read out by the fine-rank heads, which predict the fine-rank targets. Unlike pre-rank, fine-rank predicts many more objectives; we focus on the core ones, CTR and CVR, denoted $\mathcal{T}_f$, again with a binary cross-entropy loss $\mathcal{L}_{f}$,
\begin{equation}
\mathcal{L}_{f} = -\sum_{t\in\mathcal{T}_f}\left[y_f^t \log \hat{y}_f^t+(1-y_f^t)\log(1-\hat{y}_f^t)\right],
\label{eq:loss-f}
\end{equation}
where $y_f^t$ and $\hat{y}_f^t$ are the label and the predicted probability of target $t$ as in Eq.~\ref{eq:loss-p}. 

\paratitle{Knowledge Distillation.} Sharing the user context gives pre-rank and fine-rank a common representation, but each stage still learns its own scoring function, and the two can order the same candidates differently. This is problematic, as a candidate that pre-rank discards can never be recovered by fine-rank, so every disagreement between the two stages permanently removes items that fine-rank would have ranked at the top. To address this, we adopt knowledge distillation~\cite{kd} between the two ranking stages, with the fine-rank task as the teacher and the pre-rank task as the student. The stage with the richer candidate features thus supervises the ordering produced by the cheaper one. We follow the mean-centering step of logit standardization~\cite{logitstd} but replace the standard deviation by a fixed temperature $\tau$. For each pre-rank target $t\in\mathcal{T}_p$, the teacher logit $o^{T}_{t}$ and the student logit $o^{S}_{t}$ are centered by their respective means $\mu^{T}_{t}$ and $\mu^{S}_{t}$, and the resulting student probability is trained against the teacher probability with a binary cross-entropy loss $\mathcal{L}_{kd}$ summed over the targets,
\begin{equation}
\begin{gathered}
p^{T}_{t}=\sigma\!\left(\frac{o^{T}_{t}-\mu^{T}_{t}}{\tau}\right),
\qquad
p^{S}_{t}=\sigma\!\left(\frac{o^{S}_{t}-\mu^{S}_{t}}{\tau}\right),\\
\mathcal{L}_{kd}=-\sum_{t\in\mathcal{T}_p}\Big[\,p^{T}_{t}\log p^{S}_{t}+(1-p^{T}_{t})\log(1-p^{S}_{t})\Big],
\end{gathered}
\label{eq:loss-distill}
\end{equation}
where $\sigma$ is the sigmoid function. To prevent the distillation loss from affecting the teacher, we detach the teacher logit, allowing it to be shaped only by the fine-rank objective while the alignment moves the student alone. Because of the one-model design of \Model, this distillation requires neither a separately trained teacher nor an offline distillation stage. We optimize $\mathcal{L}_{kd}$ together with the task losses in the same backward pass, so the student keeps improving as the teacher itself grows stronger.

\subsubsection{Overall Objective}

\Model is optimized via $\mathcal{L}$, a weighted sum of the three task objectives and the distillation loss $\mathcal{L}_{kd}$,
\begin{equation}
\mathcal{L} = \lambda_r\mathcal{L}_{r} + \lambda_p\mathcal{L}_{p} + \lambda_f\mathcal{L}_{f} + \lambda_{kd}\mathcal{L}_{kd},
\label{eq:loss-total}
\end{equation}
where $\lambda_r$, $\lambda_p$, $\lambda_f$, and $\lambda_{kd}$ control the relative contributions of the four objectives.

\subsection{Scaling and Stabilization}
\label{ssec:scaling}

Scaling \Model requires increasing model capacity while bounding activated computation and maintaining effective attention and stable optimization. To this end, we combine parameter-efficient scaling, stabilization, and optimization techniques, and revisit which input designs remain beneficial at scale.

\paratitle{TransBlock Design.} Figure~\ref{fig:model}(b) expands the internal structure of a TransBlock. Each block follows the pre-norm layout, in which an RMSNorm~\cite{rmsnorm} precedes the attention and FFN sub-layers and a residual connection follows each. The branch output is scaled by a residual multiplier $\gamma=1/\sqrt{2N}$ for $N$ blocks, so the residual stream stays bounded as the model deepens. In the attention sub-layer, queries and keys are RMS-normalized per head (QKNorm~\cite{qknorm}), and gated attention~\cite{qiu2025gatedattentionlargelanguage} modulates the attention output before the output projection. The FFN sub-layer is a \emph{sparse mixture-of-experts} (MoE), which routes each token to a small subset of experts. This grows the parameter count while keeping the activated compute per token roughly constant. Algorithm~\ref{alg:transblock} summarizes the forward pass of a block, and we detail each component below.
\begin{algorithm}[H]
  \caption{TransBlock forward pass}
  \label{alg:transblock}
  \begin{algorithmic}[1]
    \Require hidden states $X$, stage visibility mask $\mathcal{M}$, residual multiplier $\gamma=1/\sqrt{2N}$
    \State $\tilde{X} \gets \mathrm{RMSNorm}(X)$ \Comment{pre-norm}
    \State $Q, K, V, G \gets \mathrm{split}(\tilde{X}\,W^{QKVG})$ \Comment{one fused projection}
    \State $\bar{Q}, \bar{K} \gets \mathrm{RMSNorm}_{\mathrm{head}}(Q),\ \mathrm{RMSNorm}_{\mathrm{head}}(K)$ \Comment{QKNorm}
    \State $A \gets \mathrm{Attn}(\bar{Q}, \bar{K}, V;\ \mathcal{M})$
    \State $X \gets X + \gamma\,\big(A \odot \sigma(G)\big)\,W^O$ \Comment{gated attention}
    \State $\tilde{X} \gets \mathrm{RMSNorm}(X)$ \Comment{pre-norm}
    \State $X \gets X + \gamma\,\mathrm{MoE}(\tilde{X})$ \Comment{sparse MoE}
    \State \Return $X$
  \end{algorithmic}
\end{algorithm}

\paratitle{Grouped-Query Attention.} On the attention side, we adopt grouped-query attention (GQA)~\cite{gqa}. The query heads are partitioned into groups, and every group shares a single key/value head, so a layer keeps far fewer key/value heads than query heads. The saving matters because the key/value cache of the shared behavior sequence is computed once and then read by the stage-specific tokens of all three tasks. During retrieval it is further broadcast across the beams of the decoder. Its size therefore sets both the memory footprint and the memory traffic of every cross-attention step. GQA shrinks this cache in proportion to the group size without loss in quality (\S\ref{ssec:ablation}). Only the number of key/value heads and the head dimension must agree between the stage-specific tokens and the shared behavior sequence. The two are therefore free to use different numbers of query heads over the same cached keys and values.

\paratitle{Sparse MoE.} Following DeepSeekMoE~\cite{deepseekmoe}, the FFN in \Model is a sparse MoE built from \emph{fine-grained experts}, SwiGLU networks~\cite{swiglu} with a narrow intermediate width. One \emph{shared expert} is always active, and the rest form a set of routed experts. For each input, a lightweight router produces one score per routed expert, and a fixed number of the highest-scoring routed experts are selected. Their scores are renormalized to sum to one and multiplied by a fixed routed-expert scaling factor. The sub-layer output is the shared expert's output plus the weighted sum of the selected routed experts. Inspired by DeepSeek-V3~\cite{deepseek_v3}, we adopt a sigmoid router rather than a softmax one. A softmax couples the experts, since raising the score of one necessarily lowers all the others. This pushes the router toward winner-take-all routing and ties the scale of every score to the total number of experts. A sigmoid scores each expert's relevance to the input independently and on a fixed scale. Several experts can then be judged relevant at once, and the scores remain comparable as experts are added, which eases training at larger expert counts.

\paratitle{Gated Attention.}
We adopt gated attention~\cite{qiu2025gatedattentionlargelanguage}, which applies a head-specific, input-dependent sigmoid gate to the attention output. The gate logits are produced from the normalized attention input by an extra slice of the fused QKVG projection. The softmax-weighted values of every attention head are multiplied element-wise by the sigmoid of these gate logits before the output projection. The gate therefore controls what each head writes back to the residual stream rather than the attention distribution itself. Sitting between the value and output projections, it adds a non-linearity that keeps the two from collapsing into a single linear map. It also yields input-dependent sparsity, so a head that is uninformative for the current token can be switched off instead of placing its attention mass on a sink position. This suppresses attention sinks and massive activations and improves training stability.

\paratitle{Stabilization Techniques.}
As the model grows in width and depth, training becomes harder. The scales of the activations, gradients, and per-step weight updates all drift with model size, so a configuration that is stable at one size can diverge at the next. Each scaling step would then call for a fresh hyperparameter search, at a scale where every trial is expensive. We therefore parameterize the backbone so that its training dynamics stay invariant as the model grows, following the maximal update parameterization (\textbf{$\mu$P})~\cite{mup} for width and \textbf{Depth-$\mu$P}~\cite{depthmup} for depth. Both rest on one principle. Scaling changes how many terms each layer sums. The initialization, the multipliers, and the learning rate must therefore be chosen together. The goal is that every hidden activation and its per-step change stay at a scale that neither grows nor vanishes with model size, which we write as $\Theta(1)$. For width, consider a hidden layer $W\mathbf{x}$ with input $\mathbf{x}\in\mathbb{R}^{d}$ and weight matrix $W$, where $d$ is the width. Fan-in initialization, \ie entries of $W$ with variance $1/d$, already keeps $W\mathbf{x}$ at $\Theta(1)$, since its $d$ terms are uncorrelated at initialization. An Adam step with learning rate $\eta$, however, changes every entry of $W$ by roughly $\eta$. The resulting change $\Delta W\,\mathbf{x}$ sums $d$ terms that are correlated with $\mathbf{x}$, so it grows as $\Theta(\eta d)$. $\mu$P cancels this growth by scaling the Adam learning rate of every hidden weight by $1/d$. For depth, let $N$ be the number of blocks, each adding an attention branch and an FFN branch to the residual stream. The stream sums $2N$ branch outputs that are nearly uncorrelated at initialization, so a $1/\sqrt{2N}$ residual multiplier keeps the forward signal at $\Theta(1)$. The updates of different branches, however, are correlated through the shared loss and add up rather than cancel, so the total update grows as $N$. The residual multiplier already cancels one factor of $\sqrt{N}$ of this growth, and under Adam the learning rate must be scaled by $1/\sqrt{N}$ to cancel the other. Table~\ref{tab:mup-hidden} lists the resulting rules for the hidden layers relative to a base model of width $d_0$ and depth $N_0$, in terms of the width multiplier $m_d=d/d_0$ and the depth multiplier $m_N=N/N_0$.

Two further quantities drift during training rather than with model size. The attention logits grow with the norms of the query and key projections, so we apply \textbf{QKNorm}~\cite{qknorm} to the queries and keys. This holds the logits in a fixed range and keeps the softmax from collapsing onto a single position. The weight norms themselves grow under Adam, and the effective learning rate shrinks with them. We therefore optimize the dense parameters with \textbf{AdamW}, whose decoupled \textbf{weight decay}~\cite{weightdecay} holds the norms at a stable level.

\paratitle{Multimodal Embedding.}
Capacity also changes how an item's multimodal embedding should enter the model. The established practice, \emph{SimTier}~\cite{sim_tier}, does not feed the multimodal embedding itself. For every behavior sequence, it computes the cosine similarity between the candidate and each item of that sequence, summarizes those scores into a histogram, and feeds the histogram to the model as a dense feature. \Model instead feeds the multimodal embedding directly, as side information on every behavior item and on the candidate. This choice is beneficial only with sufficient capacity, as \S\ref{ssec:multimodal} shows.

\begin{table}[t]
  \centering
  \small
  \begin{tabular}{llll}
    \toprule
    Hidden-layer quantity & Base model & Width rule ($\mu$P) & Depth rule (Depth-$\mu$P) \\
    \midrule
    Initialization variance of $W$ & $1/d_0$ & $\times\,1/m_d$ (\ie $1/d$) & unchanged \\
    Residual multiplier $\gamma$ & $1/\sqrt{2N_0}$ & unchanged & $\times\,1/\sqrt{m_N}$ (\ie $1/\sqrt{2N}$) \\
    Adam learning rate & $\eta_0$ & $\times\,1/m_d$ & $\times\,1/\sqrt{m_N}$ \\
    \bottomrule
  \end{tabular}
  \caption{$\mu$P and Depth-$\mu$P rules for the hidden weights of the backbone, relative to a base model of width $d_0$ and depth $N_0$. The two rules compose multiplicatively, \eg the hidden learning rate becomes $\eta_0/(m_d\sqrt{m_N})$.}
  \label{tab:mup-hidden}
\end{table}

\subsection{Sequence-Native Training}
\label{ssec:snt}

A larger backbone makes every training step more expensive, and the largest part of that cost is encoding the long behavior sequence. Conventional recommendation training pays this cost many times over, because each stage builds its own training samples, one per exposure or one per request, and every sample carries its own copy of the sequence. A user with a hundred exposures therefore has that sequence copied and encoded up to a hundred times. 
Methods such as RLB~\cite{rlb} and MTGR~\cite{mtgr} reduce this redundancy by grouping exposures around a shared user context.
SNT extends this reuse across the cascade. The sequence is encoded once, and every exposure reads from that encoding, whichever request and whichever stage it comes from.

Two properties make this possible. First, the behavior sequence is append-only and is encoded with causal attention~\cite{onetrans}. 
The hidden state at each position therefore summarizes only the behaviors up to that position, and the encoding of the whole sequence contains the encoding of every earlier prefix unchanged. 
Second, each exposure keeps its own request identity, timestamp, task identity, stage-specific features, and labels, but it no longer carries a copy of the behavior sequence. 
In the model, an exposure appears as its stage-specific tokens attached to the shared encoding under the stage visibility mask, and it reads only the prefix up to its request time, as the next paragraph describes. To make this sharing work in practice, SNT also changes how the training samples are grouped. Exposures are grouped by user, so that the exposures of one user within a time window (a \emph{user window}) are stored and read together with one copy of that user's behavior sequence. The length of that window is a configuration choice (\S\ref{ssec:efficiency}).

\begin{figure}[t]
  \centering
  \includegraphics[width=0.95\linewidth]{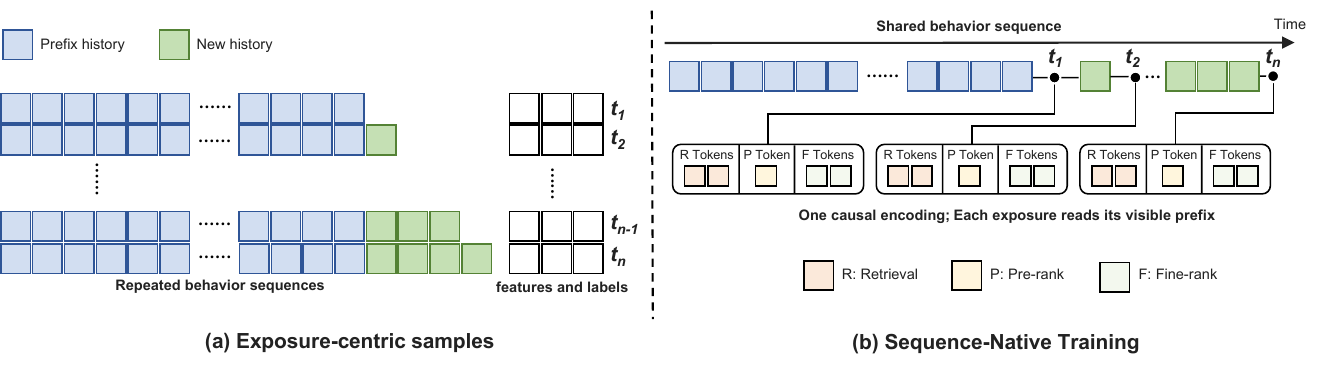}
    \caption{
    Sample organization in SNT.
    (a) Exposure-centric samples are stored separately for every exposure at every cascade stage, each with its own copy of the behavior sequence.
    (b) SNT stores and causally encodes the behavior sequence once per user window. Each exposure keeps its stage-specific features and labels and reads only the prefix available at its request time.
    }
    \Description{
    Panel (a) shows separate retrieval, pre-rank, and fine-rank samples for two requests, each pairing stage-specific features and labels with a copy of the available behavior sequence.
    Panel (b) shows one shared behavior sequence with the same stage-specific records attached at the two request boundaries. Solid-outline blocks denote repeated history; dashed-outline blocks denote newly appended behaviors. The later request sees a longer prefix, while the earlier request's prefix remains unchanged.
    }
  \label{fig:snt}
\end{figure}

\paratitle{What Each Exposure Sees.}
Formally, exposure $e$ is anchored at the position $a_e$ of the last behavior in $\mathcal{S}_u$ that had arrived by its request time (\S\ref{ssec:one-transformer}). Under the stage visibility mask it attends only to the prefix up to $a_e$. Exposures from the same request share one anchor. Each exposure therefore sees exactly what the model saw online at that request, even though the sequence was encoded once for all of them. We call this the causal view of an exposure, and Figure~\ref{fig:snt} illustrates it for two requests.

The views stay valid as the sequence grows because it is append-only. Let $\mathcal{S}_u^{(n)}$ denote the sequence seen by the $n$-th request. The sequence seen by the next request is
\begin{equation}
\mathcal{S}_u^{(n+1)}
=
\mathcal{S}_u^{(n)}
\,\Vert\,
\Delta\mathcal{S}_u^{(n+1)},
\label{eq:snt-append}
\end{equation}
where $\Vert$ denotes concatenation and $\Delta\mathcal{S}_u^{(n+1)}$ contains the behaviors that arrive between the two requests. Because new behaviors are appended and never inserted into an earlier prefix, the causal states and anchors defined on $\mathcal{S}_u^{(n)}$ remain valid in $\mathcal{S}_u^{(n+1)}$.

\paratitle{Request-Relative Time Encoding.}
The anchor decides which behaviors an exposure can see, but not how old they are. A behavior from yesterday matters differently to a request made today than to one made a month later. Yet both requests read the same encoded sequence. We therefore make the attention between stage-specific tokens and the behavior sequence depend on the time difference between the request and the behavior. We use a timestamp-based variant~\cite{cadet} of rotary position encoding~\cite{roformer}. The query of an exposure at request time $t_e$ is rotated by $t_e$, and the key of a behavior at event time $t_i$ is rotated by $t_i$. For the temporal channels, the attention term then satisfies
\begin{equation}
\left(
R(t_e)\mathbf q_e^{\mathrm{time}}
\right)^\top
\left(
R(t_i)\mathbf k_i^{\mathrm{time}}
\right)
=
(\mathbf q_e^{\mathrm{time}})^\top
R(t_i-t_e)
\mathbf k_i^{\mathrm{time}},
\label{eq:relative-time}
\end{equation}
where $R(t)$ denotes the rotary transformation for timestamp $t$. The term depends only on $t_i-t_e$, not on where the behavior sits in the sequence. Tokens of the same exposure share one request time, so their relative rotation cancels. Behavior self-attention is left unchanged. We use multiple fixed temporal periods to capture both short- and long-term dynamics. The rotation of a key depends only on the behavior's own timestamp. The rotated keys are therefore computed once and reused by every request and stage, and each exposure contributes only its own query and anchor.

Compared with exposure-centric samples, SNT yields a $4.4\times$ training speedup under the same hardware budget (\S\ref{ssec:efficiency}).

\section{Serving and System Optimization}
\label{sec:serving}

\subsection{Serving Pipeline}
\label{ssec:pipeline}

Although \Model is jointly trained as one Transformer, serving still follows the original retrieval $\to$ pre-rank $\to$ fine-rank cascade, with each stage executing only its own candidate-side computation.

\paratitle{Shared User Context.}
For an online request from user $u$, the behavior sequence $\mathcal{S}_u$ is first processed by the shared causal Transformer to produce a reusable user-context cache $C_u$. The same $C_u$ is then consumed by retrieval, pre-rank, and fine-rank, so the long behavior sequence is encoded once per request rather than once per stage.
Retrieval consumes $C_u$ together with the retrieval-specific features $\mathcal{NS}_r$ and performs \dgr decoding, applying the business offset of \S\ref{sssec:dig} to the prefix scores, followed by beam search over the SID hierarchy.
Pre-rank and fine-rank are non-autoregressive candidate-scoring passes over the same $C_u$.
Pre-rank uses the lightweight feature set $\mathcal{NS}_p$ to score the retrieved candidates, while fine-rank applies the richer $\mathcal{NS}_f$ only to the surviving candidates.
The candidate set therefore continues to shrink through the cascade even though the user context is shared.

\subsection{Decoding and Kernel Optimization}
\label{ssec:opt}

\paratitle{Retrieval Decoding Optimization.}
Retrieval first generates the decision prefix and then decodes the SID codes with beam search.
Both are decoded step by step, so their latency is set by the number of sequential steps and the work per step. We reduce the former for the prefix and the latter for SID decoding.
For the decision prefix, every dimension has a small vocabulary. There are three values for the purchase level $z_{oc}$, two for the supply type $z_{ad}$, and a few each for the discovery level $z_{disc}$ and the spending level $z_{aov}$. Their product is small enough to score in a single forward pass.
Rather than decoding the parallel and chained positions one after another, we therefore enumerate all valid combinations of $\mathbf{z}$ in Eq.~\ref{eq:dig-expand} and score them in one forward pass over the cached user context. This collapses the prefix from multiple sequential steps to a single scoring step, which is where the business offset acts.
For SID decoding, write $k$ for the beam width and $K$ for the number of items the search finally returns to pre-rank. We replace the global top-$k$ over all beam--token pairs with a two-stage top-$k$, which first retains the $k$ best extensions of each beam and then selects the global top-$k$ among them.
This yields exactly the same top-$k$ set. The intermediate candidate pool shrinks from the beam width times the vocabulary size to $k$ per beam, which cuts the work per step accordingly.

\paratitle{Kernel Fusion.} Beyond the decoding changes above, a set of kernel-level optimizations raises the model FLOPs utilization (MFU) of every forward pass in both training and serving. On modern GPUs, utilization is governed by the size and number of GEMMs and by the memory traffic of the element-wise operations between them. We therefore make the GEMMs larger and fewer and fuse whatever lies in between. The attention between the stage-specific tokens and the shared behavior sequence runs on a FlashAttention-style kernel~\cite{flashattention} whose built-in stage visibility mask covers all candidates of a user in one call, without padding or materializing the attention matrix. Within a TransBlock, the query, key, value, and gate projections are issued as one bias-free general matrix multiplication (GEMM) whose output is consumed in place. QKNorm is applied by an RMSNorm kernel that addresses the query and key slices through offsets rather than copies. The gate and up projections of SwiGLU are likewise merged into one GEMM. Each fused kernel is written to exploit the efficient instructions of modern GPUs. For example, the RMSNorm kernel guarantees that every token's hidden vector starts at a 16-byte-aligned address, so the compiler can emit 128-bit vectorized loads that fetch eight 16-bit values per instruction.

\paratitle{Low-Precision Serving.} Unlike LLMs, recommendation models consume many floating-point features whose value ranges and precision must be preserved, so serving runs in FP16 by default. The Transformer backbone itself is more tolerant, since the RMSNorm before every sub-layer and on the queries and keys keeps its activations on a bounded scale. We therefore push the SwiGLU GEMMs, which account for about two thirds of the FLOPs of a TransBlock, further down to FP8. Serving also admits one fusion that training cannot. With no backward pass that needs the intermediate activations, the merged gate-and-up GEMM, the down GEMM, and the scaled residual update of SwiGLU run as a single kernel.
\section{Experimental Results}\label{sec:exp}

\subsection{Experimental Setup}\label{ssec:setup}

\paratitle{Dataset.} We evaluate \Model offline on production logs from a large-scale industrial recommendation system. The logs are collected with the same logging, feature-snapshot, and labeling pipeline as OneTrans~\cite{onetrans} but over a different time window, so absolute numbers are not comparable with those reported there. The production logs are kept in chronological order, and every feature is recorded at exposure time to avoid temporal leakage and to keep offline evaluation consistent with online serving. Labels such as clicks and orders are accumulated over the fixed windows used in production. All logs are collected in compliance with strict privacy policies, with personally identifiable information anonymized and hashed.

\paratitle{Tasks and Metrics.} We evaluate all three stages. Retrieval is generative and decodes SID codes with beam search rather than scoring a fixed candidate set. Following UniSGR~\cite{unisgr}, we measure it with HitRate@$M$ (HR@$M$), the fraction of requests whose ground-truth item has its SID among the top-$M$ SIDs that beam search returns, and report $M{=}1$ and $M{=}10$. Pre-rank and fine-rank score a given candidate set, and we report AUC and UAUC (exposure-weighted user-level AUC) for both CTR and CVR. Both stages and their baselines are trained and evaluated on the same exposures and labels, without unexposed candidates or sampled negatives.

Following the next-batch protocol of OneTrans~\cite{onetrans}, we stream the data chronologically and score each mini-batch before it updates the model, so no prediction is made on data the model has already trained on. Metrics are computed daily and macro-averaged over the evaluation period. We also report serving \emph{GFLOPs}, the floating-point cost that a model spends on one request under the deployed configuration, summed over the stages it serves.\footnote{OneTrans~\cite{onetrans} reported training FLOPs. Here we report serving FLOPs instead, since the cost of serving a request is what bounds the QPS and latency of the deployed cascade.} This is the cost actually executed \textbf{after the optimizations described in OneTrans}. For example, the candidates that a stage scores for a request share one encoding of the behavior sequence, so the encoding is counted once rather than once per candidate.

\paratitle{Implementation.} We evaluate \Model (\S\ref{sec:model}) at two scales. Our default, \textbf{\modelname\textsubscript{L}} (L for large), stacks the TransBlocks of \S\ref{ssec:scaling} and costs $662$ GFLOPs per request. \textbf{\modelname\textsubscript{S}} (S for small) is a lighter variant of the same design that costs $239$ GFLOPs per request, about $36\%$ of \modelname\textsubscript{L}. For the retrieval task, we quantize each item into three SID codes with a codebook of $8192$ entries per level. The decision prefix covers four dimensions ($z_{oc}$, $z_{disc}$, $z_{ad}$, $z_{aov}$) laid out over two token positions.

During training, we use a dual-optimizer strategy. Sparse embeddings are optimized with AdaGrad~\cite{adagrad}, with a learning rate of $0.1$ and an initial accumulator value of $1.0$. Dense parameters are optimized with AdamW~\cite{weightdecay}\footnote{Weight decay follows the convention of modern LLM training, \ie a decoupled shrinkage $W \leftarrow W - \eta\omega W$ with weight decay coefficient $\omega$, scaled by the learning rate $\eta$ rather than a learning-rate-independent step $W \leftarrow W - \omega W$, as in TensorFlow's legacy AdamW.} (Adam $\beta_1{=}0$, $\beta_2{=}0.99999$, $\epsilon{=}10^{-5}$, weight decay $0.01$). With $\beta_1{=}0$ this optimizer carries no momentum and reduces to RMSProp with bias correction of the second moment and decoupled weight decay. We keep the name AdamW for the implementation it uses. Beyond the stabilization techniques of \S\ref{ssec:scaling}, we use global gradient-norm clipping~\cite{pascanu2013difficulty} for both dense and sparse layers, and we train in BF16 with a per-GPU batch size of $2048$ on 16 GPUs. In the training objective (Eq.~\ref{eq:loss-total}), the retrieval, pre-rank, and fine-rank losses carry weights $\lambda_r{=}0.1$, $\lambda_p{=}1$, and $\lambda_f{=}1$, and the distillation loss carries $\lambda_{kd}{=}1$ at a temperature of $\tau{=}0.5$. The centering means of Eq.~\ref{eq:loss-distill} are maintained as exponential moving averages over training batches with momentum $0.9999$.

\paratitle{Baselines.} We compare \Model against the following baselines, all trained with the same features:
\begin{itemize}
\item \textbf{OneTrans-GR} is our deployed generative retrieval (GR) model, a standalone SID-only model without the decision prefix. It represents each item by a three-level SID (\S\ref{sssec:dig}) and autoregressively decodes its codes with a Transformer decoder. It is the most recently developed GR model in our system, built after OneTrans~\cite{onetrans} and at a much larger scale. This makes it a strong baseline rather than a lightly tuned reference model.
\item \textbf{OneTrans-Lite} is our deployed pre-rank model, a compact variant of OneTrans that fits the tighter compute budget of pre-rank.
\item \textbf{OneTrans}~\cite{onetrans} is our deployed fine-rank model, which jointly models a user's behavior sequences and non-sequential features within a single Transformer (\S\ref{sec:background}).
\item \textbf{UniSGR}~\cite{unisgr} and \textbf{UniPinRec}~\cite{unipinrec} are two recent frameworks that unify generative retrieval with a downstream ranking component. In our reproduction, we map UniPinRec's lightweight ranking task to pre-rank and UniSGR's multi-objective ranking task to fine-rank, and we decode UniSGR's retrieval following its original implementation, in which the task token is set externally rather than predicted from the behavior sequence. Since each couples only two stages, we compare each model on the stages represented by this mapping. UniPinRec's retrieval is the one exception, as it is not generative but retrieves by nearest-neighbor search over item embeddings, and we therefore do not compare it against the SID-based generative retrieval models reported here.
\end{itemize}

\subsection{Offline Performance}\label{ssec:main}

\begin{table*}[htbp]
\centering
\footnotesize
\setlength{\tabcolsep}{2.5pt}
\resizebox{\textwidth}{!}{%
\begin{tabular}{ll|rr|rrrr|rrrr|r}
\toprule
\multirow{3}{*}{\textbf{Type}} & \multirow{3}{*}{\textbf{Model}} & \multicolumn{2}{c|}{Retrieval} & \multicolumn{4}{c|}{Pre-rank} & \multicolumn{4}{c|}{Fine-rank} & Efficiency \\
\cmidrule(lr){3-4} \cmidrule(lr){5-8} \cmidrule(lr){9-12} \cmidrule(lr){13-13}
 & & \multirow{2}{*}{HR@1} & \multirow{2}{*}{HR@10} & \multicolumn{2}{c}{CTR} & \multicolumn{2}{c|}{CVR} & \multicolumn{2}{c}{CTR} & \multicolumn{2}{c|}{CVR} & \multirow{2}{*}{GFLOPs} \\
\cmidrule(lr){5-6} \cmidrule(lr){7-8} \cmidrule(lr){9-10} \cmidrule(lr){11-12}
 & & & & AUC & UAUC & AUC & UAUC & AUC & UAUC & AUC & UAUC & \\
\midrule
\multirow{3}{*}{\textbf{(a) Single Model}}
 & OneTrans-GR & 0.0545 & 0.3112 & N.A. & N.A. & N.A. & N.A. & N.A. & N.A. & N.A. & N.A. & 202 \\
 & OneTrans-Lite & N.A. & N.A. & 0.7965 & 0.7080 & 0.9102 & 0.7365 & N.A. & N.A. & N.A. & N.A. & 4 \\
 & OneTrans & N.A. & N.A. & N.A. & N.A. & N.A. & N.A. & 0.8084 & 0.7393 & 0.9139 & 0.7428 & 69 \\
\midrule
\multirow{4}{*}{\textbf{(b) Unified Models}}
 & UniPinRec & N.A. & N.A. & 0.8046 & 0.7212 & 0.9113 & 0.7506 & N.A. & N.A. & N.A. & N.A. & 325 \\
 & UniSGR & 0.0483 & 0.2964 & N.A. & N.A. & N.A. & N.A. & 0.7976 & 0.7209 & 0.9130 & 0.7328 & 117 \\
 & \modelname\textsubscript{S} & 0.0660 & 0.3441 & 0.8053 & 0.7372 & 0.9128 & 0.7534 & 0.8171 & 0.7532 & 0.9163 & 0.7589 & 239 \\
 & \modelname\textsubscript{L} & $\mathbf{0.0923}$ & $\mathbf{0.4053}$ & $\mathbf{0.8089}$ & $\mathbf{0.7409}$ & $\mathbf{0.9145}$ & $\mathbf{0.7596}$ & $\mathbf{0.8182}$ & $\mathbf{0.7564}$ & $\mathbf{0.9184}$ & $\mathbf{0.7682}$ & 662 \\
\bottomrule
\end{tabular}%
}
\caption{Performance across the three stages. Part (a) reports our deployed production baselines, each specialized for one stage; part (b) reports models that unify multiple stages. ``N.A.'' marks a stage that a model does not cover, as well as UniPinRec's retrieval, which we exclude because it is not generative. GFLOPs is the total arithmetic cost of one request for each model in isolation, including all candidates that stage scores, thousands for pre-rank and hundreds for fine-rank. It is the cost actually executed after the optimizations described in OneTrans~\cite{onetrans}, \eg the user-side sequence that these candidates share is encoded once rather than once per candidate. It is not a throughput proxy, since achieved throughput also depends on MFU. The best result per quality metric is shown in \textbf{bold}.}
\label{tab:main}
\end{table*}

Table~\ref{tab:main} summarizes the performance of \Model and the corresponding baselines across the three stages. The existing cascade, \ie OneTrans-GR for retrieval, OneTrans-Lite for pre-rank, and OneTrans for fine-rank, together costs $275$ GFLOPs per request. This is the reference for the comparisons below.

\modelname\textsubscript{S} serves all three stages at $239$ GFLOPs, slightly under that reference, yet surpasses each of the three baselines on the stage it serves, by $+21.10\%$ on retrieval HR@1 and $+2.17\%$ on fine-rank CVR UAUC, for example. \modelname\textsubscript{L} spends $2.4\times$ as much as the cascade and turns the extra compute into larger gains on all ten metrics. It lifts retrieval HR@1 by $69.36\%$ over OneTrans-GR, pre-rank CTR UAUC by $4.65\%$ and CVR UAUC by $3.14\%$ over OneTrans-Lite, and fine-rank CTR UAUC by $2.31\%$ and CVR UAUC by $3.42\%$ over OneTrans. Both variants also surpass UniPinRec and UniSGR, two recent models that unify multiple stages, on every stage they cover. The part of these gains that comes from joint training alone, under matched architecture and parameter budget, is isolated in \S\ref{ssec:joint}.

\subsection{Online A/B Performance}\label{ssec:ab}
We assess the business impact of \Model with online A/B tests in a large-scale industrial recommendation system. The \emph{control group} is served by the existing cascade of OneTrans-GR, OneTrans-Lite, and OneTrans. In the \emph{treatment group}, a single \modelname\textsubscript{L}, served with the stack of \S\ref{ssec:opt}, replaces all three. Traffic is split at the user level with hashing-based randomization, with $50\%$ of traffic assigned to each group.

\Model lifts GMV per user by \textbf{$9.74\%$} over the control group. It also raises user active days by \textbf{$0.25\%$}, so the items it recommends bring users back more often. These gains are measured against a strong cascade of three separately trained and separately served models, and they show the potential of cascade-level unification.

\Model also serves more efficiently. We define QPS as the number of requests completed per second through all three stages. Under the same hardware budget, the treatment group is served at $3.2\times$ the QPS of the control group, even though \Model executes more FLOPs per request. The higher QPS comes from the serving stack and from encoding the behavior sequence once per request rather than once per stage, and \S\ref{ssec:efficiency} quantifies both.

\section{Analysis}\label{sec:analysis}
In this section, we analyze what each component of \Model contributes. Unless noted otherwise, ranking results are reported on CTR, and every improvement represents a relative change against the corresponding reference configuration instead of an absolute difference in AUC or UAUC.

\subsection{Effect of Joint Training}\label{ssec:joint}
We analyze whether training the three tasks jointly outperforms training each of them separately. Throughout this subsection, the joint model and its separately trained counterparts share the same architecture, parameter budget, training data, and training objective, with the loss weights of \S\ref{ssec:setup} unchanged. This includes the distillation loss of \S\ref{sssec:ranking}, which the separately trained pre-rank receives from the separately trained fine-rank. Any gain therefore reflects joint training rather than added capacity or distillation. We first analyze merging pre-rank with fine-rank, and then analyze merging all three stages together.

\begin{wraptable}{r}{0.46\textwidth}
\vspace{-\intextsep}
\centering
\begin{tabular*}{\linewidth}{@{\extracolsep{\fill}}lcc@{}}
\toprule
Task & CTR AUC & CTR UAUC \\
\midrule
Pre-rank & +0.09\% & +0.17\% \\
Fine-rank & +0.12\% & +0.21\% \\
\bottomrule
\end{tabular*}
\caption{Improvement from jointly training pre-rank and fine-rank over training each task separately.}
\label{tab:joint-rank}
\end{wraptable}

\paratitle{Combining Pre-rank and Fine-rank.} We start with the two ranking tasks, letting their tokens read one shared user context and optimizing them together. Table~\ref{tab:joint-rank} shows that both tasks improve over their separately trained counterparts. Pre-rank gains $+0.09\%$ CTR AUC and $+0.17\%$ CTR UAUC, and fine-rank gains $+0.12\%$ CTR AUC and $+0.21\%$ CTR UAUC. The two tasks therefore reinforce each other rather than compete.

\begin{figure*}[htbp]
\centering
 \includegraphics[width=0.86\textwidth]{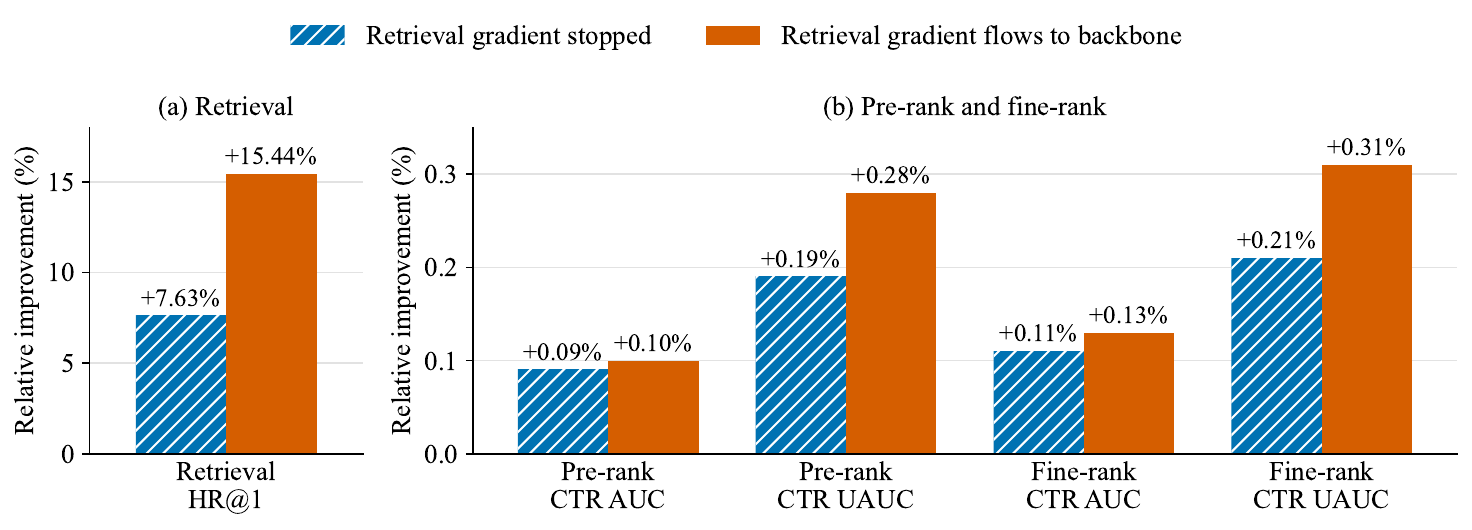}
\caption{Effect of jointly training all three tasks, before and after letting the retrieval task's gradient flow into the shared user context. (a): Improvement of HR@1 over a separately trained retrieval model of the same architecture and parameter budget. (b): Improvement of pre-rank and fine-rank CTR AUC/UAUC over training each task separately.}
\Description{Joint Training Analysis}
\label{fig:joint-full}
\end{figure*}

\paratitle{Combining All Three Stages.} We then add the retrieval task and train all three together, and Figure~\ref{fig:joint-full} compares every task against its separately trained counterpart. With the retrieval gradient into the shared user context stopped, the retrieval task\footnote{In this experiment the retrieval task of the joint model and its separately trained counterpart share the same backbone and compute budget as OneTrans-GR, so that the comparison isolates joint training from capacity.} reuses the shared user context already shaped by the ranking tasks and gains $+7.63\%$ on HR@1. Pre-rank and fine-rank gain $+0.19\%$ and $+0.21\%$ CTR UAUC.\footnote{These differ slightly from Table~\ref{tab:joint-rank} because the embedding tables remain shared, so the retrieval task still has a small effect on pre-rank and fine-rank even with its gradient into the shared user context stopped.} We then let the retrieval gradient flow into the shared user context, so that its next-token prediction objective also shapes the user context the ranking tasks read. This roughly doubles the retrieval task's gain to $+15.44\%$ and raises pre-rank and fine-rank to $+0.28\%$ and $+0.31\%$ CTR UAUC. Unifying all three tasks therefore improves every task over training them separately, with the retrieval objective further strengthening the shared user context for ranking.

\subsection{Distillation and Consistency between the Ranking Stages}\label{ssec:consistency}

\begin{figure*}[htbp]
\centering
\begin{subfigure}[b]{0.49\textwidth}
\centering
\includegraphics[width=\linewidth]{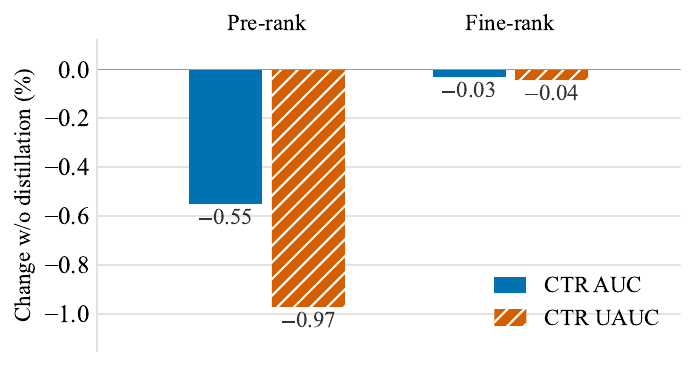}
\caption{Removing the distillation loss from \Model}
\end{subfigure}\hfill
\begin{subfigure}[b]{0.49\textwidth}
\centering
\includegraphics[width=\linewidth]{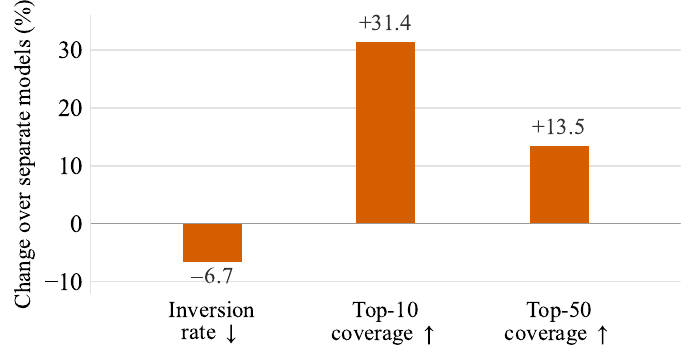}
\caption{Online consistency between pre-rank and fine-rank}
\end{subfigure}
\caption{Distillation between the two ranking tasks and their consistency. (a): Relative change in CTR AUC and CTR UAUC of each ranking task when $\mathcal{L}_{kd}$ is removed from \Model with everything else held fixed. (b): Relative change of \Model over the separately trained pre-rank and fine-rank of \S\ref{ssec:joint} on each consistency metric, where the arrow in a label gives the direction of improvement.}
\Description{Left: grouped bar chart of the AUC and UAUC change of pre-rank and fine-rank without distillation. Right: bar chart of the relative change on three consistency metrics.}
\label{fig:consistency}
\vspace{-5mm}
\end{figure*}

\paratitle{Knowledge Distillation.} The previous subsection shows that joint training helps all three tasks. For the two ranking tasks, \Model adds a second mechanism, the knowledge distillation of \S\ref{sssec:ranking}, an extra loss $\mathcal{L}_{kd}$ in which the fine-rank logits supervise the pre-rank ones. Figure~\ref{fig:consistency}(a) shows what this loss contributes. This ablation is orthogonal to the joint-training comparison, where both settings keep $\mathcal{L}_{kd}$; here the joint model drops it. Removing $\mathcal{L}_{kd}$ from \Model while holding the architecture, data, and every other hyperparameter fixed costs pre-rank $0.55\%$ CTR AUC and $0.97\%$ CTR UAUC. This shows how much distillation contributes to closing the gap between pre-rank and fine-rank. Fine-rank itself moves by at most $0.04\%$, as the detached teacher logit intends, so the student's gain is not paid for by the teacher.

\paratitle{Cross-Stage Consistency.} We further analyze how consistently pre-rank and fine-rank order the same candidates online. Both the shared user context and the distillation loss pull the two stages together, and Figure~\ref{fig:consistency}(b) measures their combined effect with two metrics, the \emph{inversion rate} and \emph{top-10 and top-50 coverage}. The inversion rate is the fraction of item pairs that the two tasks order differently, and top-10 coverage is the fraction of pre-rank's top 10 that also appears in fine-rank's top 10, and likewise for top-50. Both are computed over the candidates that pre-rank forwards to fine-rank. Compared with the separately trained pre-rank and fine-rank of \S\ref{ssec:joint}, \Model reduces the inversion rate by $6.7\%$ and raises top-10 and top-50 coverage by $31.4\%$ and $13.5\%$, all as relative changes. The two ranking stages therefore disagree less often. More importantly, pre-rank forwards far more of fine-rank's top candidates, so fewer of the items that fine-rank would place at the top are discarded before fine-rank ever sees them.

\subsection{Objective Coverage of \dgr}\label{ssec:consolidation}
\begin{figure*}[ht]
\centering
\includegraphics[width=0.6\textwidth]{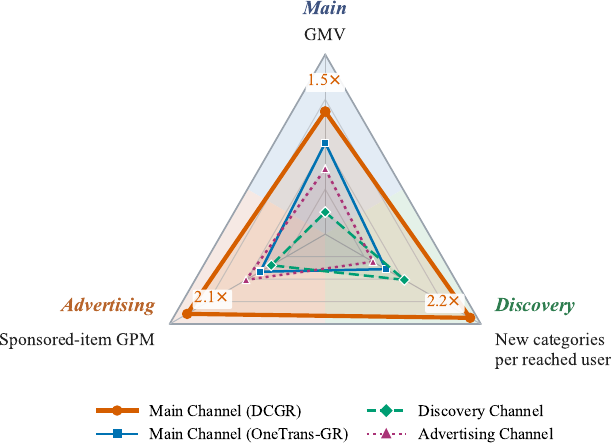}
\caption{Coverage of the three retrieval objectives. In this retrieval-stage A/B test at equal traffic, the treatment group unifies the three channels into \Model's retrieval task, shown as \dgr, while the control group keeps the three channels. All four are compared on the three objectives, one per axis. On every axis \dgr exceeds the channel built for that objective, by the multiple marked at that vertex. Each exposure is credited to exactly one channel by the retrieval-source field of the serving log, and the radius is a power scale.}
\label{fig:recall-consolidation}
\vspace{-2mm}
\end{figure*}

As described in \S\ref{sec:intro}, retrieval in our production system is split into specialized channels for different business objectives. The main channel, served by OneTrans-GR, targets clicks and orders, the discovery channel promotes categories new to the user, and the advertising channel retrieves sponsored items. \dgr represents these objectives within a single model through corresponding decision dimensions. Specifically, the purchase level $z_{oc}$ captures the main channel's click-and-order objective, the discovery level $z_{disc}$ captures the discovery objective, and the supply type $z_{ad}$ distinguishes organic from sponsored items for advertising. This moves the three objectives from separate retrieval channels into one shared decision space. We next ask whether \textbf{a single \dgr can still cover what each specialized channel was built to provide}.

We test this with an online retrieval-stage A/B test, with traffic split equally at the user level. In the treatment group, \Model's retrieval task, \ie \dgr, replaces OneTrans-GR in the main channel, and the discovery and advertising channels are switched off, so the three channels are unified by a single model. The control group keeps all three channels as deployed. Because the three channels serve different purposes, we evaluate each objective with its native metric, namely GMV for the main objective, the number of categories new to the user per reached user for discovery, and GMV per thousand exposures (GPM) on sponsored items for advertising.

Figure~\ref{fig:recall-consolidation} shows that \dgr covers all three objectives within one model. It produces $1.5\times$ the GMV of the control group's main channel, reaches $2.2\times$ as many categories new to the user per reached user as the discovery channel, and achieves $2.1\times$ the sponsored-item GPM of the advertising channel. Thus, a single \dgr matches or exceeds the specialized channel on the metric native to each objective.

\paratitle{Main Channel Analysis.} The $1.5\times$ GMV gain on the main objective comes primarily from broader exposure rather than higher value per exposure. Table~\ref{tab:recall-detail}(a) shows that \dgr increases exposures by $69.3\%$ and clicks by $74.6\%$ over the control group's main channel. Part of that volume is what the discovery and advertising channels served in the control group, since \dgr now serves it as well. Its per-exposure conversion is lower. Specifically, orders per thousand exposures decrease by $34.0\%$, and GMV per thousand exposures by $13.0\%$. The larger exposure volume nevertheless more than offsets these declines, increasing total orders by $11.4\%$ and total GMV by $48.0\%$. CTR per exposure also increases by $3.2\%$. The average price of an exposed item decreases by $4.4\%$, indicating that the additional exposures are concentrated more heavily on lower-priced items.

\begin{table}[htbp]
\centering
\small
\begin{tabular}{lr}
\toprule
\multicolumn{2}{l}{\emph{(a) Main channel, treatment against control}} \\
\midrule
Exposures                            & $+69.3\%$ \\
Clicks                               & $+74.6\%$ \\
Orders                               & $+11.4\%$ \\
GMV                                  & $+48.0\%$ \\
Click-through rate per exposure      & $+3.2\%$  \\
Orders per thousand exposures        & $-34.0\%$ \\
GMV per thousand exposures           & $-13.0\%$ \\
Average price of an exposed item     & $-4.4\%$  \\
\midrule
\multicolumn{2}{l}{\emph{(b) Against the discovery channel}} \\
\midrule
Categories new to the user, per reached user                   & $2.2\times$ \\
Share of the other channel's discovery items also reached      & $76.0\%$ / $14.0\%$ \\
Share of the other channel's user--category pairs also reached & $11.7\%$ / $1.2\%$ \\
User--category pairs formed per day                            & $9.4\times$ \\
Conversion of the users only this channel reaches              & $2.8\times$ \\
\midrule
\multicolumn{2}{l}{\emph{(c) Against the advertising channel, on sponsored items}} \\
\midrule
Exposures                            & $1.8\times$  \\
GMV per thousand exposures           & $2.1\times$ \\
Orders per exposure                  & $1.7\times$  \\
Conversion rate                      & $2.6\times$  \\
Average order value                  & $1.2\times$  \\
\bottomrule
\end{tabular}
\caption{The retrieval-stage A/B test of Figure~\ref{fig:recall-consolidation} in detail. (a) Relative changes of \dgr in the treatment group over the control group's main channel. The GMV row of (a) is the figure's main axis. (b) and (c) Ratios of \dgr to the discovery and the advertising channel, respectively. The first row of (b) and the second row of (c) are the figure's discovery and advertising axes. Each pair of percentages in (b) gives \dgr's coverage of the discovery channel, then the reverse.}
\label{tab:recall-detail}
\vspace{-2mm}
\end{table}

These measurements compare one channel against another rather than one system against another. \dgr serves in the treatment group the exposures that all three channels served in the control group, so its margin over the main channel alone is not the value it adds to the system. Figure~\ref{fig:recall-consolidation} therefore measures which objectives \dgr can cover, not how much it adds. The A/B test in \S\ref{ssec:ab} provides the latter over all traffic.

\begin{figure*}[ht]
\centering
\includegraphics[width=\textwidth]{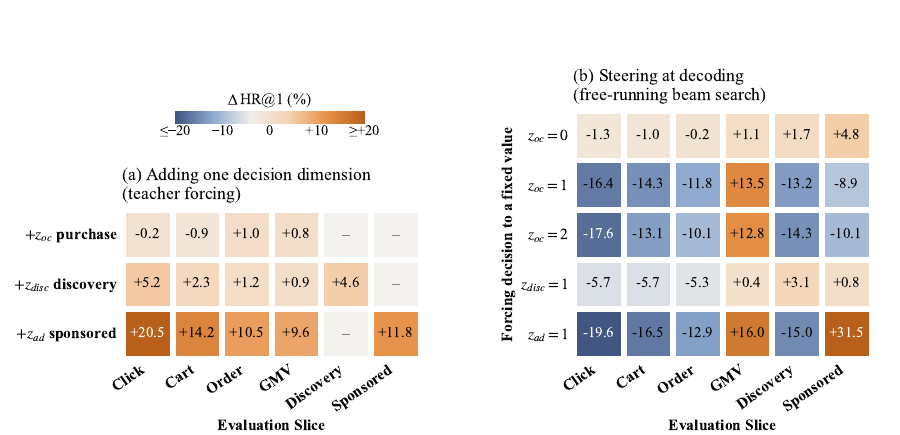}
\caption{Ablation of decision dimensions, as the relative change in HR@1 (in \%). (a): Adding one decision dimension to a SID-only generative retrieval model, measured under teacher forcing on the evaluation slice named in each column; a dash marks a slice we did not measure for that model. (b): Forcing one decision dimension to a value at decoding time, with the rest of the prefix decoded autoregressively, relative to decoding the whole prefix that way. Cells carry the exact value; the color scale is clipped at $\pm20$.}
\label{fig:decision-ablation}
\end{figure*}

\paratitle{Discovery and Advertising Analysis.} We next examine the discovery and advertising results in more detail. Table~\ref{tab:recall-detail}(b,~c) breaks down \dgr's comparison with the corresponding specialized channels.

For discovery, \dgr reaches $4.81$ categories new to the user per reached user, compared with $2.15$ for the specialized discovery channel, yielding the $2.2\times$ gain in Figure~\ref{fig:recall-consolidation}. More importantly, the difference lies primarily in which users receive these categories rather than in which items are retrieved. \dgr covers $76.0\%$ of the discovery channel's items but only $11.7\%$ of its user--category pairs; in the reverse direction, the corresponding coverages are $14.0\%$ and $1.2\%$. \dgr also forms $9.4\times$ as many user--category pairs per day, and users reached only by \dgr convert $2.8\times$ as efficiently as those reached only by the discovery channel. The two retrievers therefore draw on substantially overlapping items but expose them to very different users.

For advertising, \dgr also takes on a substantially larger role in sponsored-item retrieval. Sponsored items account for $33.7\%$ of its exposures, compared with $4.7\%$ for the main channel in the control group. On sponsored items, \dgr serves $1.8\times$ as many exposures as the specialized advertising channel and achieves $2.1\times$ its GPM. This gain comes from stronger downstream conversion and value rather than clicks. \dgr achieves $1.7\times$ the orders per exposure, $2.6\times$ the conversion rate, and $1.2\times$ the average order value, despite a click-through rate about one third lower. Sponsored items retrieved by \dgr are therefore clicked less often but purchased more often and at higher value.

Together, these results show that a single \dgr can cover the objectives represented by $z_{oc}$, $z_{disc}$, and $z_{ad}$, matching or exceeding the corresponding specialized channel on its native metric. This demonstrates the potential to consolidate these retrieval objectives within one model.

\subsection{\dgr Decision Dimensions Analysis}\label{ssec:decision-ablation}

The previous subsection shows that a single \dgr retriever can cover multiple specialized objectives. We next ask whether the decision prefix itself carries useful retrieval signals and whether it controls item generation.

\begin{table}[t]
\centering
\begin{tabular}{lrrrr}
\toprule
Intervention on $\mathbf{z}$ & CTR & Orders & AOV & GMV \\
\midrule
None (predicted prefix)                     & --        & --         & --         & -- \\
Force $z_{oc}{=}0$ (click)                  & $+8.2\%$  & $-10.4\%$  & $+5.8\%$   & $-5.1\%$ \\
Force $z_{aov}$ to the highest value        & $-8.1\%$  & $-33.5\%$  & $+38.9\%$  & $-7.7\%$ \\
Business offset, weak $\beta$               & $+2.2\%$  & $+4.9\%$   & $+2.5\%$   & $+7.5\%$ \\
Business offset, medium $\beta$             & $-0.9\%$  & $-3.0\%$   & $+13.9\%$  & $+10.4\%$ \\
Business offset, strong $\beta$             & $-5.4\%$  & $-1.2\%$   & $+28.3\%$  & $+26.7\%$ \\
\bottomrule
\end{tabular}
\caption{Steering the same trained model at decoding time in an online retrieval-stage A/B test. Values are relative changes against decoding with the predicted prefix, over the exposures credited to \Model's retrieval task as in \S\ref{ssec:consolidation}. Orders counts main orders, one per checkout however many sub-orders it splits into, and AOV is GMV per main order. Exposure counts were logged only for the row that forces $z_{aov}$ to its highest value, so the table has no exposure column. In that row, the exposures credited to \Model's retrieval task fall by $19.4\%$.}
\label{tab:steering}
\end{table}

\paratitle{Decision Dimensions Carry Retrieval Signal.} Figure~\ref{fig:decision-ablation}(a) adds one decision dimension at a time to a SID-only GR model and supplies its ground-truth value through teacher forcing. We measure the resulting change in HR@1 across click, cart, order, GMV, discovery, and sponsored-item slices. Because the supplied decision describes the interaction being predicted, this experiment measures the information provided by a correct decision rather than deployable end-to-end retrieval accuracy.

We also discover these dimensions carry different amounts of item information. The purchase level $z_{oc}$ changes no slice by more than $1\%$, whereas $z_{disc}$ improves all measured slices and $z_{ad}$ produces the largest gains, ranging from $9.6\%$ to $20.5\%$. This difference reflects how strongly each decision constrains item identity. For example, $z_{ad}$ restricts a sponsored interaction to the sponsored subset, which accounts for only $28\%$ of interactions, whereas $z_{oc}$ says relatively little about which item was chosen. The gains also extend beyond the corresponding objective. In particular, $z_{ad}$ improves click, cart, order, and GMV retrieval in addition to sponsored-item retrieval. No measured slice decreases by more than $1\%$.

\paratitle{Decision Dimensions Control Generation.} Teacher forcing shows that a correct decision is informative, but serving relies on decisions predicted by the model. We therefore intervene on one predicted dimension during free-running beam search. If the SID decoder ignored the prefix, forcing a decision would have little effect. Figure~\ref{fig:decision-ablation}(b) shows the opposite. Forcing $z_{ad}$ to sponsored raises sponsored-item HR@1 by $31.5\%$ and GMV HR@1 by $16.0\%$, while reducing click HR@1 by $19.6\%$. Forcing a multi-order $z_{oc}$ raises GMV HR@1 by $12.8\%$ but reduces click HR@1 by $17.6\%$. Forcing $z_{disc}$ toward discovery raises its own slice by $3.1\%$. Changing a decision therefore changes the retrieved items in the expected direction.

The intervention size also depends on how often the forced value already occurs. Sponsored items account for $28\%$ of interactions, whereas interactions in a category new to the user already make up roughly three quarters, leaving fewer requests for discovery forcing to change. The small offline discovery movement also does not contradict its larger online gain in \S\ref{ssec:consolidation}, nor does the slight drop in click HR@1 under $z_{oc}{=}0$ contradict its CTR gain online in Table~\ref{tab:steering}. Offline HR@1 rewards reproducing the logged item, whereas online metrics reward exposures that users click or that reach valuable categories new to the user, even when the retrieved item differs from the log.

\subsection{\dgr Decision Steering and Prefix Design}\label{ssec:prefix-design}
We next examine how the decision dimensions should be steered and organized in practice.

\paratitle{Predicted vs. Supplied Decisions.} A natural alternative to predicting the decision prefix is to supply the desired decision directly, as objective-conditioned methods do~\cite{unisgr}. Table~\ref{tab:steering} compares this hard conditioning with \dgr's business offset in an online retrieval-stage A/B test. We either force $z_{oc}=0$ toward clicks, force $z_{aov}$ to its highest value, or apply the offset at increasing strengths. Hard forcing approximates the extreme of objective steering, whereas the offset gradually tilts the model's predicted distribution.

Hard forcing behaves much like an objective-specific channel. Table \ref{tab:steering} shows that forcing $z_{oc}=0$ raises CTR by $8.2\%$ but lowers GMV by $5.1\%$, while forcing the highest $z_{aov}$ value raises AOV by $38.9\%$ but lowers orders by $33.5\%$ and GMV by $7.7\%$. The business offset instead provides a gradual trade-off. As its strength increases from weak to strong, GMV rises from $+7.5\%$ to $+26.7\%$, while CTR moves from $+2.2\%$ to $-5.4\%$. The offset therefore steers the model toward a business objective without replacing its predicted decision with a fixed value.

Table~\ref{tab:offset-online} presents a second retrieval-stage A/B test, which confirms the same trade-off at the system level. Unlike Table~\ref{tab:steering}, it measures all exposures per user rather than only those credited to \dgr. Steering $z_{oc}$ alone increases orders by $2.4\%$ but reduces GMV by $0.6\%$. Adding $z_{aov}$ trades some orders for higher AOV, turning the GMV change positive and reaching $+3.0\%$ at the stronger setting. The two dimensions therefore control complementary aspects of business value.

\begin{table}[ht]
\centering
\begin{tabular}{lrrrr}
\toprule
Offset applied & Clicks & Orders & AOV & GMV \\
\midrule
Purchase level $z_{oc}$ only               & $+8.8\%$  & $+2.4\%$ & $-2.9\%$  & $-0.6\%$ \\
\quad $+$ spending level $z_{aov}$, weak   & $+11.4\%$ & $-5.2\%$ & $+5.9\%$  & $+0.4\%$ \\
\quad $+$ spending level $z_{aov}$, strong & $+5.1\%$  & $-9.4\%$ & $+13.7\%$ & $+3.0\%$ \\
\bottomrule
\end{tabular}
\caption{A second retrieval-stage A/B test of the business offset, as relative changes per user against the control group. Unlike Table~\ref{tab:steering}, the metrics cover all of a user's exposures, not only those credited to \dgr, so they show the offset's effect on the whole system. The last two rows keep the purchase-level offset unchanged and add a spending-level offset.}
\label{tab:offset-online}
\end{table}

\paratitle{Organizing the Decision Prefix.} The complementary roles of $z_{oc}$ and $z_{aov}$ motivate modeling purchase quantity and value separately. Replacing them with a single GMV token reduces transaction-value-weighted HR@1 by $9\%$. A separate purchase indicator is also redundant because $z_{oc}=0$ already represents a non-converting interaction, and adding one leaves HR@1 unchanged.

Independent decision dimensions should not be chained arbitrarily because chaining requires choosing a decoding order. Table~\ref{tab:permutation} tests this effect on an earlier configuration with three binary dimensions for order, cold-start, and discovery. The order objective performs best when decoded first, and the order-first configurations also achieve the highest weighted HR@128. The decoding order therefore biases which objective is favored. \dgr avoids this bias by combining independent dimensions into the parallel decision token $\texttt{DT}^{p}$ and chaining only genuinely dependent decisions, with $z_{aov}$ following $z_{oc}$ in $\texttt{DT}^{d}$. This design also shortens the prefix, but its main advantage is avoiding ordering bias, since the summed dimensions share one position and have no decoding order.

\begin{table}[htbp]
\centering
\begin{tabular}{lcccc}
\toprule
Position of the order dimension & Weighted & Order & Cold-start & Discovery \\
\midrule
First  (\texttt{acd}, \texttt{adc})  & $0.3287$ & $0.4590$ & $0.0352$ & $0.1437$ \\
Second (\texttt{cad}, \texttt{dac})  & $0.3267$ & $0.4548$ & $0.0348$ & $0.1437$ \\
Third  (\texttt{cda}, \texttt{dca})  & $0.3252$ & $0.4520$ & $0.0371$ & $0.1446$ \\
\bottomrule
\end{tabular}
\caption{Effect of the decoding order of chained decision tokens, on an earlier configuration with three binary dimensions (\texttt{a} = order, \texttt{c} = cold-start, \texttt{d} = discovery). HR@128 with the prefix decoded autoregressively, averaged over the two permutations that place the order dimension at each position; the weighted HR@128 column weights hits by their transaction value.}
\label{tab:permutation}
\end{table}

\subsection{Architecture Analysis}\label{ssec:ablation}
\begin{figure*}[htbp]
\centering
\includegraphics[width=0.92\textwidth]{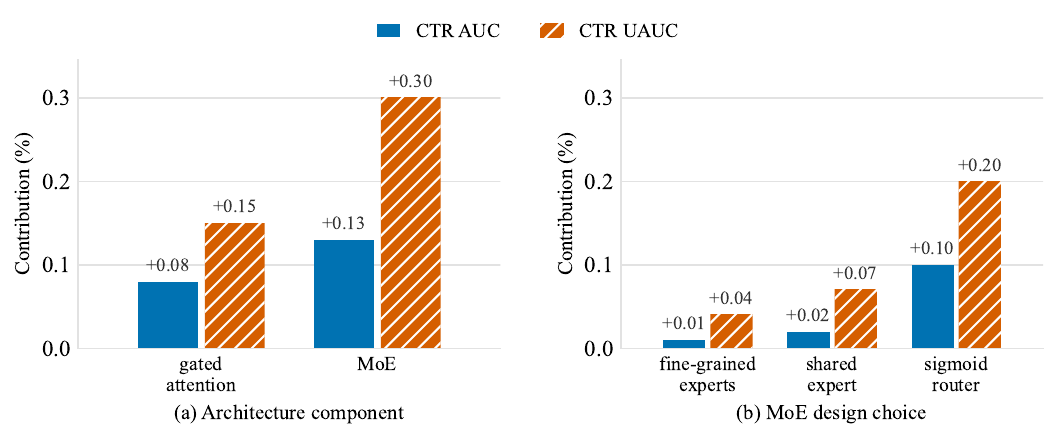}
\caption{Contribution of each architecture component and each MoE design choice, measured on the fine-rank task. Each bar is the relative improvement in CTR AUC or CTR UAUC that a single component or design choice brings over its conventional alternative, with everything else in \modelname\textsubscript{L} held fixed. The bars are therefore separate single-component ablations rather than a decomposition of the MoE gain. (a): The architecture components of \S\ref{ssec:scaling}. (b): The design choices inside the MoE block.}
\label{fig:ablation}
\end{figure*}

We now analyze the architecture components introduced in \S\ref{ssec:scaling}, namely GQA, gated attention, and MoE. Figure~\ref{fig:ablation} reports what gated attention and MoE contribute to \modelname\textsubscript{L} on the fine-rank task, and then what each design choice inside the MoE block contributes.

Replacing full multi-head attention with GQA leaves fine-rank quality effectively unchanged, so GQA's smaller key/value cache costs no accuracy. Figure~\ref{fig:ablation}(a) further shows that gated attention contributes $0.08\%$ CTR AUC and $0.15\%$ CTR UAUC over the conventional attention it replaces.

MoE contributes more than gated attention, gaining $0.13\%$ CTR AUC and $0.30\%$ CTR UAUC over a compute-matched dense FFN. Figure~\ref{fig:ablation}(b) looks inside the MoE block. Fine-grained experts contribute $0.01\%$ CTR AUC and $0.04\%$ CTR UAUC over fewer, wider experts of the same total parameter count, and the shared expert contributes $0.02\%$ and $0.07\%$ over the same MoE without a shared expert. The sigmoid router contributes more than either, with $0.10\%$ CTR AUC and $0.20\%$ CTR UAUC over a softmax router.\footnote{Repeated runs of the same configuration differ by about $0.006\%$ in CTR AUC and $0.012\%$ in CTR UAUC (\S\ref{ssec:stability}), so a difference larger than this should be read as meaningful.}

\begin{figure*}[htbp]
\centering
\includegraphics[width=0.86\textwidth]{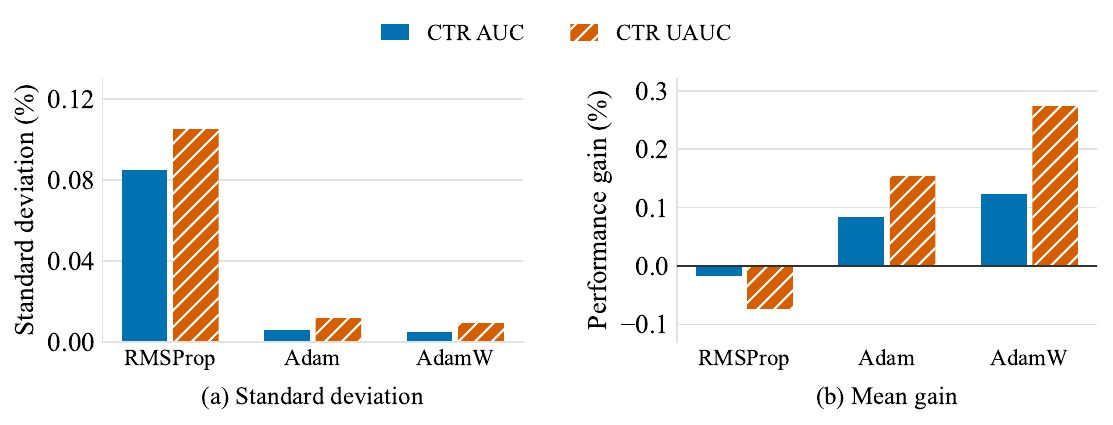}
\caption{Performance gain from widening \Model, measured over three runs. (a): Standard deviation across different runs. (b): Averaged gain across different runs.}
\label{fig:stability}
\end{figure*}

\subsection{Stabilization Analysis}\label{ssec:stability}
We analyze the effectiveness of our stabilization methods introduced in \S\ref{ssec:scaling}. Each experiment below removes one technique from \Model and measures how much the same widening of the model gains without and with that technique, each against a narrow model built the same way. The widening gain therefore differs from paragraph to paragraph because the reference differs, so gains are comparable within one figure but not across figures. The gain of widening the full configuration is the one reported for AdamW in Figure~\ref{fig:stability}. 

\paratitle{AdamW and Weight Decay.} RMSProp~\cite{rmsprop} is the optimizer commonly used to train industrial recommendation models, and the RMSProp baseline here uses the learning rate, $\epsilon$, and decay coefficient tuned for our deployed models. As Figure~\ref{fig:stability} shows, it yields a large standard deviation across repeated runs, and widening the model brings no gain. We then replace it with Adam~\cite{adam} under the same learning rate and $\epsilon$, with $\beta_2$ set to the RMSProp decay coefficient and $\beta_1{=}0$ (\S\ref{ssec:setup}) so that no momentum is added. The only difference between the two optimizers is then the bias correction of the second-moment estimate. This correction prevents the inflated effective learning rate of the early updates. It cuts the standard deviation of CTR AUC from $0.085\%$ to $0.006\%$ and of CTR UAUC from $0.105\%$ to $0.012\%$, both estimated from the same three runs. Widening then becomes effective, turning the negative outcome under RMSProp into a gain of $0.15\%$ CTR UAUC. Adding decoupled weight decay on top, \ie AdamW, raises that gain further to nearly $0.3\%$ CTR UAUC while the standard deviation stays at the same low level.

\begin{figure*}[htbp]
\centering
\includegraphics[width=0.92\textwidth]{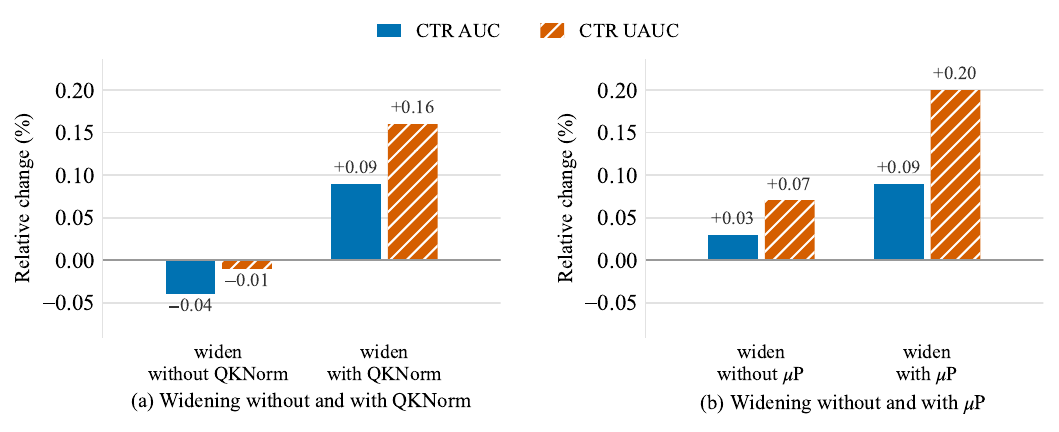}
\caption{Effect of QKNorm and $\mu$P when widening \Model, measured on the fine-rank task. In each panel, both bars are measured against a narrow model built without that technique, which for $\mu$P means standard parameterization with its own initialization and learning rate. Bars are therefore comparable within a panel but not across panels or with Figure~\ref{fig:stability}. (a): Widening without and with QKNorm. (b): Widening without and with $\mu$P.}
\label{fig:qknorm-mup}
\end{figure*}

\paratitle{QKNorm.} Figure~\ref{fig:qknorm-mup}(a) shows that widening the model without QKNorm falls short of the expected gain, changing fine-rank by $-0.04\%$ CTR AUC and $-0.01\%$ CTR UAUC. Once QKNorm is introduced, the same widening gains $+0.09\%$ CTR AUC and $+0.16\%$ CTR UAUC.

\paratitle{$\mu$P.} Figure~\ref{fig:qknorm-mup}(b) shows that widening the model without $\mu$P, keeping the hyperparameters tuned at the original width, brings only a slight improvement of $+0.03\%$ CTR AUC and $+0.07\%$ CTR UAUC. Under standard parameterization these hyperparameters are no longer the right ones once the model is widened, so every widening would call for a new search. $\mu$P instead keeps the optimal hyperparameters invariant to width, \ie hyperparameter transfer~\cite{mup}. The same widening then yields $+0.09\%$ CTR AUC and $+0.20\%$ CTR UAUC without any retuning.

\begin{wraptable}{r}{0.46\textwidth}
\vspace{-\intextsep}
\centering
\small
\begin{tabular*}{\linewidth}{@{\extracolsep{\fill}}lcc@{}}
\toprule
Setting & CTR AUC & CTR UAUC \\
\midrule
Base depth                        & --        & -- \\
\quad deepened, w/o Depth-$\mu$P & $+0.04\%$ & $+0.05\%$ \\
\quad deepened, w/ Depth-$\mu$P  & $+0.08\%$ & $+0.11\%$ \\
\bottomrule
\end{tabular*}
\caption{Effect of Depth-$\mu$P when deepening \Model on the fine-rank task, relative to the base depth.}
\label{tab:residual}
\end{wraptable}

\paratitle{Depth-$\mu$P.} The experiments above widen the model; deepening it is governed by Depth-$\mu$P (\S\ref{ssec:scaling}), which scales both the residual multiplier and the learning rate by $1/\sqrt{m_N}$, and we test it here. Table~\ref{tab:residual} shows that adding blocks to \Model gains only $0.04\%$ CTR AUC and $0.05\%$ CTR UAUC on its own, whereas the same deepening under Depth-$\mu$P gains $0.08\%$ and $0.11\%$. Depth-$\mu$P therefore turns additional depth into a larger gain.

\subsection{Multimodal Embedding and Model Capacity}\label{ssec:multimodal}
We verify the multimodal design of \S\ref{ssec:scaling} by comparing SimTier against the multimodal embedding at two model scales. One is our default \modelname\textsubscript{L}; the other is a much smaller variant, \modelname\textsubscript{XS} (XS for extra small), which is roughly $46\times$ cheaper to train.

\begin{figure*}[htbp]
\centering
\includegraphics[width=0.92\textwidth]{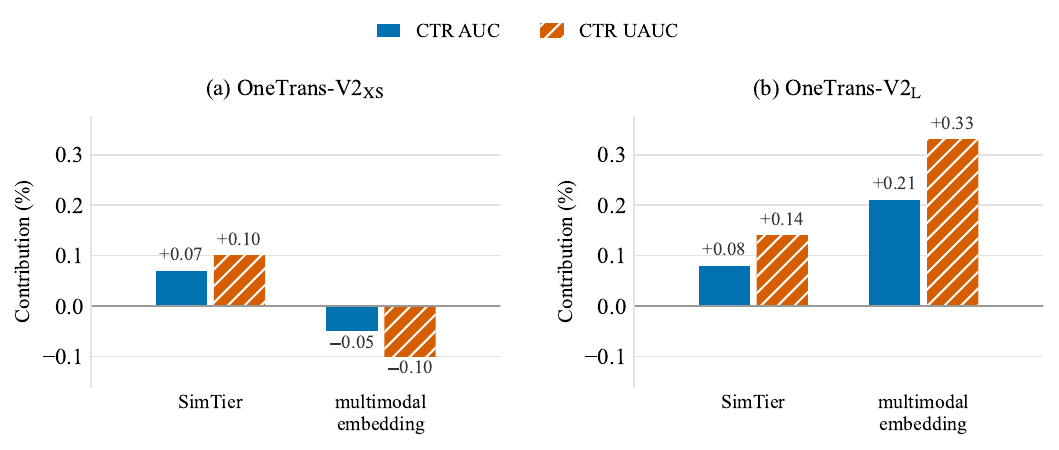}
\caption{Contribution of each way of injecting the multimodal embedding, measured at two model scales on the fine-rank task. Each bar is the relative change from adding that input alone to a model that receives no multimodal embedding, so the two bars within a panel are separate experiments rather than a decomposition. (a): \modelname\textsubscript{XS}. (b): \modelname\textsubscript{L}.}
\label{fig:multimodal}
\end{figure*}

As shown in Figure~\ref{fig:multimodal}, the two scales call for opposite designs. At the smaller scale, only SimTier helps, contributing $0.07\%$ CTR AUC and $0.10\%$ CTR UAUC. The multimodal embedding contributes $-0.05\%$ and $-0.10\%$, so feeding the embedding directly hurts a model of this size. At the larger scale both ways help, and the multimodal embedding becomes the more valuable choice, contributing $0.21\%$ against $0.08\%$ CTR AUC and $0.33\%$ against $0.14\%$ CTR UAUC without any precomputed feature. The multimodal embedding and the model's own learned item embeddings live in different spaces, and mapping one onto the other is something the model must learn. We hypothesize that a small model lacks the capacity to learn this mapping, so the similarity has to be precomputed for it. \modelname\textsubscript{L} learns the mapping on its own and gains more from the multimodal embedding than from a precomputed summary of it. How the multimodal embedding should enter a ranking model is therefore not a fixed design choice but a function of capacity, and a practice established at one scale need not be the best one at the next.

\subsection{Training and Inference Speedup}\label{ssec:efficiency}
We first analyze the training speedup brought by SNT (\S\ref{ssec:snt}), and then the inference speedup brought by the serving optimizations of \S\ref{ssec:opt}, first those specific to retrieval decoding and then those that apply to the whole model.

\begin{figure*}[htbp]
\centering
\includegraphics[width=0.92\textwidth]{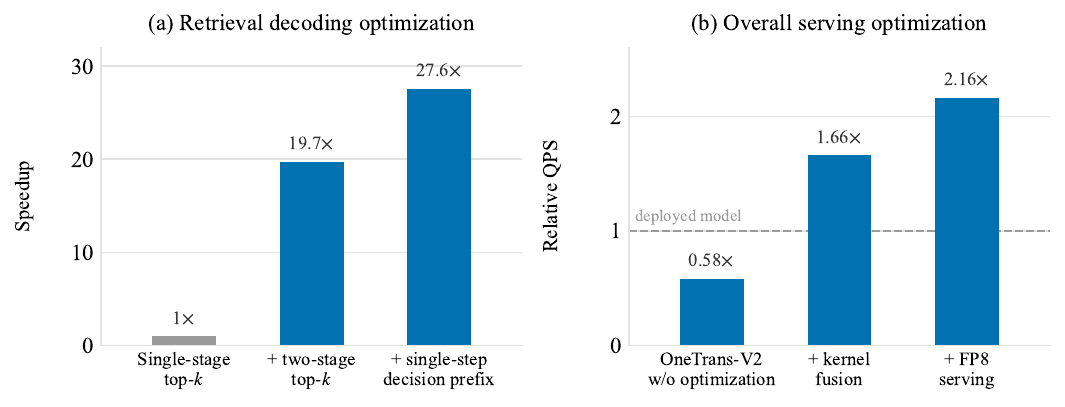}
\caption{Speedup from the serving-side optimizations. (a): Decoding speedup of \Model's retrieval task, measured as throughput under the same latency budget, with the two decoding optimizations of \S\ref{ssec:opt} added cumulatively. (b): Serving throughput of the fine-rank task relative to the deployed fine-rank model under the same latency budget, with kernel fusion and FP8 serving added cumulatively.}
\label{fig:speedup}
\end{figure*}

\paratitle{Sequence-Native Training.} We compare SNT against the exposure-centric samples of Figure~\ref{fig:snt}(a), in which every exposure is a separate sample that encodes its own copy of the behavior sequence. Sharing one encoding across all exposures of a user window yields a $4.4\times$ training speedup under the same hardware budget. In these experiments a mini-batch is organized by user windows, each holding the exposures of one user within a window of time. We cut the windows at day boundaries, so that no window straddles an evaluation date and the daily metrics of \S\ref{ssec:setup} stay comparable with those of the baselines. This choice caps the speedup, since a window can share the sequence encoding only among the exposures of one day. Windows cut by week or by month raise the speedup further, and we leave that configuration to future work. SNT also saves storage. Each user's behavior sequence is stored once instead of once per exposure, which cuts sequence feature storage by $95\%$. The exposures of a user are stored next to each other, which makes their non-sequential features compress better and cuts that storage by half.

\paratitle{Retrieval Decoding Optimization.} We further analyze the two decoding optimizations on the retrieval task. For SID decoding, Figure~\ref{fig:speedup}(a) shows that the two-stage top-$k$ raises throughput by $19.7\times$ under the online latency budget, against our own single-stage implementation rather than the deployed retrieval baseline. Per request, the same change cuts beam-search latency by $2.3\times$. The throughput gain exceeds the latency gain because throughput is measured at the largest batch that fits the latency budget. Shrinking the candidate pool from the beam width times the vocabulary size to $k$ per beam lets each GPU serve a much larger batch within that budget. Optimizing this step matters because, before this change, $94\%$ of the retrieval task's inference time was spent in the top-$k$. Scoring the decision prefix in a single step then removes the sequential prefix steps, so each remaining step carries a larger batch and maps better onto the GPU. This cuts the measured latency by about $30\%$ against decoding the prefix step by step and raises the cumulative decoding speedup to $27.6\times$.

\paratitle{Overall Serving Optimization.} The decoding optimizations above are specific to retrieval, whereas kernel fusion and FP8 serving apply to the whole model. Figure~\ref{fig:speedup}(b) measures them on the fine-rank task, relative to the deployed fine-rank model under the same latency budget. Without them the unified model serves at $0.58\times$ the throughput of the deployed model, since its fine-rank task spends more compute per request. Kernel fusion raises this to $1.66\times$ and FP8 serving to $2.16\times$, so the unified model serves the fine-rank stage $2.16\times$ faster than the specialized model it replaces despite its larger compute. Sharing one user context across the three stages (\S\ref{ssec:pipeline}) encodes the behavior sequence once per request rather than once per stage, which adds a further $1.53\times$ on top of all per-stage optimizations, measured over the whole cascade. The two together yield the end-to-end $3.2\times$ QPS of \S\ref{ssec:ab}.

\section{Conclusion}
We present \Model, one Transformer that serves retrieval, pre-rank, and fine-rank in an industrial recommendation cascade system. \Model builds on three key ideas. First, it unifies retrieval, pre-rank, and fine-rank within one Transformer model by sharing the candidate-independent user context across stages while preserving each stage's native candidate features and computation. Second, \dgr consolidates objective-specific retrieval models by representing business objectives as decisions within a single generative process. Third, SNT reorganizes training around each user's lifelong behavior sequence and amortizes sequence encoding across exposures. Together, these designs unify the cascade across stages and retrieval objectives while making the resulting shared backbone practical to train at production scale. This unification enables computation reuse, joint learning, and in-model knowledge distillation. It also pools previously fragmented capacity into a shared backbone. We scale this backbone with sparse MoE to increase capacity with bounded activated computation, and use \(\mu\)P-style parameterization to stabilize scaling. In a large-scale industrial deployment, \Model improves GMV by \(9.74\%\) and, together with a co-designed serving stack, delivers \(3.2\times\) QPS of the cascade it replaces under the same hardware budget. We believe \Model provides a production-scale foundation for the broader transition from model-per-stage recommendation toward a unified one-model recommendation architecture.

\clearpage

\bibliographystyle{acl_natbib}
\bibliography{references}

\appendix

\section{Full Author List}\label{app:authors}

\textbf{Foundation Team:} Hannan Cao, Jun Guo, Haolei Pei, Zhaoqi Zhang, Tianyu Wang, Ziyang Wang, Youchen Sun, Yue Xue, Yucheng Mao, Lintao Yan, Yufei Feng, Shaowei Liu*

\textbf{Architecture Team:} Rongkun Xing, Feiling Gong, Xinyu Chenli, Cong Xu, Mingge Zhang, Yunjia Zhu, Yajing Zhang, Pengfei Ren, Yue Lin

*: Corresponding author, liushaowei.nphard@bytedance.com

\end{document}